\documentclass[notitlepage,floatfix,nofootinbib,british,balancelastpage,preprintnumbers]{revtex4-1}
\usepackage[]{natbib}
\usepackage{amsmath,mathrsfs}
\usepackage{amssymb}
\usepackage{appendix}
\usepackage{balance}
\usepackage{booktabs}
\usepackage{capt-of}
\usepackage{cleveref}

\usepackage{fancyvrb}
\usepackage[pdftex]{graphicx}
\usepackage{hyperref}
\usepackage{listings}
\usepackage{multirow}
\usepackage{relsize}
\usepackage{rotate}
\usepackage{rotating}
\usepackage{slashed}
\usepackage{subcaption}
\usepackage{tabularx}
\usepackage{ wasysym }

\usepackage{units}
\usepackage{url}
\usepackage[usenames,dvipsnames]{xcolor}\usepackage{changes}

\hypersetup{
    colorlinks=false,
    pdfborder={0 0 0},
}

\newcommand{\etmiss}{$\slashed{E}_T$}
\newcommand{\Root}{\textsc{Root}}
\newcommand{\Checkmate}{Check\textsc{mate}}
\newcommand{\CLs}{CL$_\text{S}$\ }
\newcommand{\pT}{p_\text{T}}

\newcommand{\userinputcolor}{\color{gray}{}}

\crefname{appsec}{Appendix}{Appendices}

\makeatletter
\g@addto@macro\@verbatim\scriptsize
\makeatother

\begin{document}
\fvset{samepage=true, fontsize=\scriptsize}
\begin{center}
\includegraphics[width=0.7\textwidth]{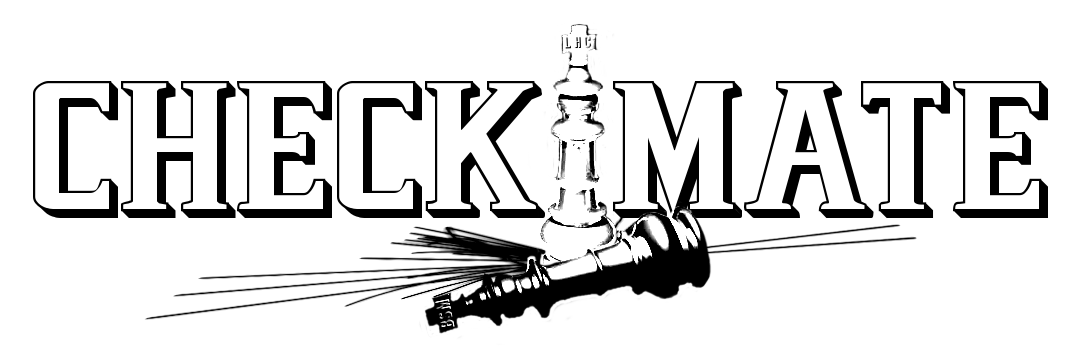}
\vspace*{1cm}
\end{center}
\title{A framework to create customised LHC analyses within CheckMATE}%

\author{Jong Soo \surname{Kim}}%
\email{jong.kim@csic.es}
\affiliation{Instituto de F\'{\i}sica Te\'{o}rica UAM/CSIC, Madrid, 
  Spain}

\author{Daniel Schmeier }%
\email{daschm@th.physik.uni-bonn.de}
\affiliation{Physikalisches
  Institut and Bethe
  Center for Theoretical Physics, University of Bonn, Bonn, Germany}
  
\author{Jamie Tattersall }%
\email{j.tattersall@thphys.uni-heidelberg.de}
\affiliation{Institut f\"ur Theoretische Physik, University of Heidelberg, Heidelberg, Germany}

\author{Krzysztof Rolbiecki}%
\email{rolbiecki.krzysztof@csic.es}
\affiliation{Instituto de F\'{\i}sica Te\'{o}rica UAM/CSIC, Madrid, 
  Spain}
\affiliation{Instytut Fizyki Teoretycznej, Uniwersytet Warszawski, Warsaw, Poland}

\begin{abstract}
\Checkmate\ is a framework that allows the user to conveniently test simulated BSM physics events against current LHC 
data in order to derive exclusion limits. For this purpose, the data runs through a detector simulation and is then 
processed by a user chosen selection of experimental analyses. These analyses are all defined by signal regions that can 
be compared to the experimental data with a multitude of statistical tools. 

Due to the large and continuously growing number of experimental analyses available, users may quickly 
find themselves in the situation that the study they are particularly interested in has not (yet) been 
implemented officially into the \Checkmate\ framework. However, the code includes a rather 
simple framework to allow users to add new analyses on their own. This 
document serves as a guide to this.

In addition, \Checkmate\ serves as a powerful tool for testing and implementing new search strategies. To aid
this process, many tools are included to allow a rapid prototyping of new analyses.
\newline

Website: \url{http://checkmate.hepforge.org/}
\end{abstract}

\keywords{analysis, detector simulation, delphes, ATLAS, CMS, tool}
\preprint{IFT-UAM/CSIC-15-021}

\setcounter{tocdepth}{1}
\setcounter{secnumdepth}{2}
\date{\today}%

\maketitle
\makeatletter
  \def\l@subsubsection#1#2{}%
\makeatother 
\clearpage

\pagebreak
\tableofcontents
\vspace*{3cm}


{\bf PROGRAM SUMMARY}


\begin{small}
\noindent                         \\
{\em Program Title:} CheckMATE, AnalysisManager                                          \\
{\em Journal Reference:}                                      \\
{\em Catalogue identifier:}                                   \\
{\em Licensing provisions:} none                                   \\
{\em Programming language:} C++, Python                                   \\
{\em Computer:} PC, Mac                                               \\
{\em Operating system:} Linux, Mac OS                                       \\
{\em Keywords:} Analysis, Confidence Limits, Monte Carlo, Detector Simulation, Delphes, ROOT, LHC  \\
{\em Classification:} 11.9                                         \\
{\em External routines/libraries:} ROOT, Python                \\
{\em Subprograms used:} Delphes                                \\
{\em Nature of problem:} The LHC has 
delivered a wealth of new data that is now being analysed. Both ATLAS and CMS have performed many searches for new physics that 
theorists are eager to test their model against. However, tuning the detector simulations, 
understanding the particular analysis details and interpreting the results can be a tedious and repetitive task. 
Furthermore, new analyses are being constantly published  by the experiments and might be not yet included in the official CheckMATE distribution.
\\
{\em Solution method:} The AnalysisManager within the CheckMATE framework allows the user to easily include new experimental analyses as they are published by the collaborations. 
Furthermore, completely novel analyses can be designed and added by the user in order to test models at higher center-of-mass energy and/or luminosity. \\
%
{\em Restrictions:} Only a subset of available experimental results have been implemented. \\ 
{\em Running time:} The running time scales about linearly with the number of input events provided by the user. The detector simulation / analysis of 20000 events needs about 50s / 1s for a single core calculation on an Intel Core i5-3470 with 3.2 GHz and 8 GB RAM.\\
\end{small}
\pagebreak

\section*{Important Note}

\begin{itemize}
 \item \Checkmate{} is built upon the tools and hard work of many people. If \Checkmate{} is used in your 
    publication it is extremely important that all of the following citations are included,
\begin{itemize}
  \item Delphes 3 \cite{deFavereau:2013fsa}.
  \item FastJet \cite{Cacciari:2011ma,Cacciari:2005hq}.
  \item Anti-$k_t$ jet algorithm \cite{Cacciari:2008gp}.
  \item \CLs prescription \cite{0954-3899-28-10-313}.
  \item In analyses that use the $M_{T2}$ family of kinematical discriminants we use the Oxbridge Kinetics 
  Library \cite{Lester:1999tx,Barr:2003rg} and the algorithm developed by Cheng and Han \cite{Cheng:2008hk} which also
  includes the $M_{T2}^{b\ell}$ variable \cite{Bai:2012gs}.
  \item In analyses that use the $M_{CT}$ family of kinematical discriminants we use MctLib \cite{Tovey:2008ui,Polesello:2009rn} which also
  includes the $M_{CT_{\bot}}$ and $M_{CT_{\lVert}}$ variables \cite{Matchev:2009ad}.
  \item All experimental analyses that were used to set limits in the study.
  \item The Monte Carlo event generator that was used.
\end{itemize}


\end{itemize}

\section{Introduction}

\Checkmate~\cite{Drees:2013wra} is a tool that allows for the easy testing of new physics models against current LHC data. The fully validated
version now contains 14 LHC analyses with over 100 separate signal regions. More analyses are being added all the time 
and the beta-version now contains over 30 finished analyses that have been validated over significant regions of the
parameter space. Including these analyses means that \Checkmate\ contains over 300 separate signal regions and this
list is growing continuously.

The tool is growing in popularity and many studies have now used \Checkmate\ to constrain both supersymmetry
(SUSY) (e.g.~\cite{Cao:2013mqa,Calibbi:2014lga,Kaminska:2014wia}) and other models of new physics (e.g.~\cite{Boehm:2014hva,Holdom:2014rsa}) that 
may be probed at the LHC. In addition, the tool can be used to fit excesses that could potentially be the first signs of new 
physics (e.g.~\cite{Kim:2014eva}). 

However \Checkmate{} is not only a tool to be used to re-interpret analyses that are already included in the program, it
is also a powerful analysis environment for users to include other analyses or invent their own searches \cite{Kim:2014noa}. To this end,
\Checkmate{} contains an \emph{AnalysisManager} which makes the process of writing, testing and validating an analysis as easy 
as possible. In particular, the manager guides the user through the various choices of available final state objects,
automatically writes an analysis template and calculates the various statistical measures required to define the 
sensitivity of a search. Many steps along this path are also significantly smoothed by extra features included in the 
\Checkmate{} package.

Whilst similar analysis tools already exist, \Checkmate\ is specifically designed for new physics searches at the LHC. The 
analysis framework is heavily influenced by the Rivet \cite{Buckley:2010ar} program and shares the philosophy of using
projections to define the particular final state objects of interest. However, the big difference is that whilst Rivet is designed
to recast unfolded high energy analyses, \Checkmate\ includes the detector simulation Delphes \cite{deFavereau:2013fsa} with a
significantly improved ATLAS tuning.\footnote{The implementation of an improved CMS tuning is planned in the near future.} Consequently, \Checkmate\ can reproduce new physics searches where a detector 
simulation is vital to accurately reproduce the experimental result. Additionally, the integrated statistics functionality
allows exclusion and/or discovery regions to be easily found with minimal user input.

The MadAnalysis \cite{Conte:2012fm,Conte:2014zja,Dumont:2014tja} framework is another similar approach that also has an 
emphasis on new physics searches at the LHC. Since MadAnalysis also uses the Delphes detector simulation, work has now begun
to improve the compatibility with \Checkmate\ so that the tools can be used interchangeably.

For budding LHC phenomenology analysis authors, the most attractive feature is the simplification \Checkmate\ achieves in
the analysis writing procedure. Many different tunes of final state objects are available (e.g.\ `loose', `medium' and `tight' electrons in case of ATLAS) while
some objects even have the full multi-variate discriminant parametrised (e.g.\ $b$-jets) that automatically recalculate the fake rate
depending on the efficiency required. Routines also exist for the automatic calculation of isolation variables where the user only has
to specify the required parameters. Furthermore, many of the kinematical variables (e.g.\ $m_{T2}$, $\alpha_T$, razor)
used by the LHC collaborations are included. Additionally, since \Root\ is fully integrated, the \texttt{TLorentzVector} class is
available which allows the easy access and calculation of a huge number of 4-vector variables. 

To help with creating, testing and validating the analyses, tools are included for the automatic calculation of cutflows and 
signal regions. The statistical evaluations required to assess the analysis are also implemented to assist the user at every point.
Finally, if backgrounds and the corresponding errors are known, the experimental reach of the analysis to a particular model of
new physics is also then calculated.

In this manual, we will guide the users through the list of steps necessary to add these skeletons to \Checkmate, how these 
skeletons have to be understood and what has to be done in order to fill them with proper content. In Section~\ref{sec:running}
we introduce the main features of \Checkmate. Section~\ref{sec:features} describes in detail the AnalysisManager 
module of \Checkmate\ and presents the main 
ingredients of analyses within \Checkmate. In Section~\ref{sec:pracex} we provide a detailed discussion of 
two analyses implemented in \Checkmate\ in order
to familiarise the user with the analysis writing technique. This is followed by an example of a 
completely new analysis to test the  discovery potential of a compressed SUSY scenario at the 14 TeV LHC.
Finally we summarise is Section~\ref{sec:summary}.

\section{Overview: The AnalysisManager within CheckMATE\label{sec:running}}

In this section we briefly summarize \Checkmate{'}s main program flow and outline the procedure to add a new analysis 
to the existing framework. Detailed information on \Checkmate\ itself can be found in \cite{Drees:2013wra}.

\Checkmate\ uses simulated event files and cross sections as input. Users have to provide these and list the analyses 
they want them to be tested against. The provided files are then fully automatically processed by the following 3-step procedure:
\begin{enumerate}
\item The embedded program Delphes takes the input events and applies a fast detector simulation.
\Checkmate\ determines the settings required for Delphes based on the exact analyses chosen. This optimises the 
detector simulation by only requiring the final states objects needed for the particular analysis list.

\item The output of the simulation is quantified by applying a standalone C++ analysis program on the results. The 
final state objects of each event are tested to see if they fulfil the criteria of one or more of the signal regions 
defined within the analyses. For each of these signal regions, the total number of input events that pass the 
selection  is evaluated 
and normalised to the given input cross section and luminosity. 

\item The predictions for all signal regions are compared to the stored experimental results in terms of the observed and 
expected number of events. This set of information is statistically combined  to quantify the compatibility 
of the observation with the input model. \Checkmate\ then deduces whether the input model can be excluded or not to the 95\ \% C.L.
\end{enumerate}
In order to add a new analysis to this framework, \Checkmate\ requires various pieces of information and 
needs them stored in a well defined form at specific places in the code. The AnalysisManager offers a simple
way to provide this information automatically to \Checkmate. 
It is already included in the 
official \Checkmate\ distribution and only has to be compiled explicitly, see Section \ref{sec:pracex}. Users can simply use 
terminal commands to list, add, edit or remove analyses. In order to add an analysis, a 
sequence of questions has to be answered which collects the 
information that \Checkmate\ requires
to setup the internal workings accordingly: 
\begin{itemize}
\item Some general information (name of the author, name of the analysis, luminosity, $\sqrt{s}$, ...) is gathered first to 
define the headers and names of all the respective files and to determine the normalisation of the event numbers. 
\item The properties of final state objects determined on the detector level (jet clustering algorithm, working point 
efficiencies for jet flavour tagging, lepton isolation criteria, ...) will later define the Delphes setup 
during the actual \Checkmate{} run as well as the available objects in the analysis code.
\item Numbers (observed number of events, Standard Model expectation, ...) and names for all signal 
regions have to be defined by the user and are used in \Checkmate{'}s evaluation step. The user normally can take these directly from the corresponding experimental publication.
\end{itemize}
After this information is entered, the AnalysisManager will create all the internal files necessary to 
run the analysis within the normal \Checkmate{} framework. The user still has to define the actual 
analysis prescription in the form of a C++ code. For that purpose, a skeleton source code is 
provided which gives the user automatic access to all objects defined during the AnalysisManager 
step. Lots of additional predefined helper functions further reduce the effort to implement the 
analysis code. The call of a single function, \verb@countSignalEvent(<SR>)@, is sufficient to 
trigger the \Checkmate{} evaluation routines. In the upcoming sections we will discuss the 
definitions and features of all these objects and functions in more detail.

\section{Features of CheckMATE AnalysisManager\label{sec:features}}

As explained in the introduction, \Checkmate\ contains many novel features to make recasting or designing a new
LHC analysis as simple as possible. In this section we introduce these features in detail while later sections
give examples of how these are used in practice.

\subsection{Final state objects and functions}
\label{sec:fin_obj}

Most of the reconstruction efficiencies for final state objects available in \Checkmate\ have been re-tuned according to 
ATLAS performance studies. In addition, new functionality has been included to allow a full reproduction of
all the final state cuts currently implemented by the LHC experiments. Details of all the 
individual reconstruction efficiencies can be found in~\cite{Drees:2013wra}.

\subsubsection{Electrons}

All three of the current ATLAS electron definitions, loose, medium and tight,  are available in \Checkmate{} 
and can be called in any analysis. They are different in terms of the reconstruction efficiency and these are
implemented in terms of 2-dimensional look-up tables as a function of $p_T$ and $\eta$. A default loose isolation
condition that requires the sum of calorimeter objects in a cone size of $\Delta R =0.2$ not associated to the 
electron to be less than $0.2$ of the electron $p_T$ is always applied. If tighter and/or more complicated isolation
conditions are required please see the specific isolation subsection below. Also, in the case of ATLAS, a retuned algorithm is present to smear the reconstructed electron momentum.
For CMS,  the default Delphes tunings for electrons are used.

\subsubsection{Muons}

The majority of ATLAS analyses currently use one of two muon reconstruction algorithms, `Combined' (named `muonCombined' in \Checkmate)  and `Segment--Tagged'. The latter is usually used in
combination with the first to maximise the muon efficiency and is hence 
called `Combined+Segment--Tagged' (named `muonsCombinedPlus' in \Checkmate). Both definitions, `muonsCombined' 
and `muonsCombinedPlus', are 
implemented in \Checkmate{} as 2-dimensional look-up tables as a function of $\eta$ and $\phi$.  Again, 
in the case of ATLAS, a retuned algorithm is present that accurately smears the reconstructed muon momentum.
As for electrons, the default Delphes CMS tuning for muons is used.

\subsubsection{Jets}

Jets are reconstructed using either smeared calorimeter deposits in the case of ATLAS or particle flow
information in the case of CMS. By default,\footnote{The default setting can be changed in \texttt{write\_delphes\_modules.py} 
module, under the key \texttt{JetAlgorithm}. See the Delphes manual for more details~\cite{deFavereau:2013fsa}.} the anti-$k_T$ algorithm implemented in FastJet \cite{Cacciari:2011ma,Cacciari:2005hq}
is chosen to cluster jets and the user only has to specify the cone size ($\Delta R$) required. Further 
studies of jet-substructure are possible since the user has access to all jet-constituents, tracks and calorimeter towers.  

\subsubsection{Taus}

The tau reconstruction algorithm searches the event tree for the presence of a tau and, if the momentum 
vector determined from its visible decay products overlaps with a reconstructed jet, identifies the jet as a tau candidate. If not, it is considered a fake background candidate, for which different efficiencies are used. 
These jets are then further sub-divided into two categories, 1-prong or multi-prong depending on whether the jet contains exactly one or more reconstructed tracks.
For each category \Checkmate\ has three different reconstruction algorithms (loose, medium and tight) parametrised 
as a function of $p_T$ and $\eta$ that the user can easily select. Currently only ATLAS parametrisations are available
with CMS functions planned for a future release. For CMS analyses, the ATLAS parametrisations are used. 
It should be noted that the tau parametrisation is specifically tuned to work with the anti-$k_T$ jet algorithm
and a cone size $\Delta R$=0.4. Using different jet settings will effect the reconstruction efficiencies and are not recommended.

\subsubsection{B-jets}

The $b$-jet reconstruction function begins by first defining whether the jet overlaps with a $b$-quark. If not, it is 
checked whether there is overlap with a $c$-quark and if again there is no overlap, it is assumed that it 
only contains light quarks. All three types of jet have different probabilities to 
be tagged as a final state $b$-jet depending on the $p_T$ and $\eta$ of the jet.

A feature of \Checkmate\ is that the complete receiver operator curve (ROC) of the ATLAS multivariate 
$b$-tagger is implemented. Consequently the user can choose any particular $b$-tag reconstruction probability that
they desire and \Checkmate\ will automatically adjust the probability that c-quarks or light jets fake a true $b$-quark.
This feature is very useful since ATLAS uses widely different tagging points depending on the analysis of interest
and the associated fake rates vary by orders of magnitude. Currently only ATLAS parametrisations are available
but CMS functions are planned for a future release. For now, the tuned ATLAS parametrisations are used for 
CMS analyses with the assumption that for a given working point the deduced efficiency distributions are roughly equal.

\subsubsection{Tracks and Calorimeter Cells}

Although not routinely required for most analyses, all reconstructed tracks and calorimeter cells are 
available. This information can be used for example to develop more complicated isolation procedures 
than those already implemented (see \emph{Isolation} below) or for custom jet algorithms.
For the track reconstruction efficiency and momentum smearing we use the default Delphes parametrisation for 
both ATLAS and CMS. The same is also true for the calorimeter energy smearing.

\subsubsection{Missing Energy}  

Missing energy is reconstructed from the sum of the smeared calorimeter deposits in the case of ATLAS
or the sum of the smeared particle flow objects in the case of CMS. In addition, to effectively parametrise
additional QCD activity due to pile-up we also include an extra smearing factor on the missing energy
that has been tuned to match the ATLAS distributions, but is used for both ATLAS and CMS analyses: this smearing is done by adding a vector with uniformly random and direction magnitude which follows a Gaussian distribution with width 20~GeV centred around 0. Due to the way Delphes treats final state muons, their contribution 
has to be added separately to the missing energy, as will be shown in the following Section.

\subsubsection{Isolation}
\label{sec:Isolation}

\Checkmate{} supplies routines that automatically calculate the various isolation conditions required by the LHC 
experiments for both muons and electrons. Firstly the user can specify whether the isolation is calculated 
with respect to the calorimeter or to reconstructed tracks. In addition, the option
to calculate isolation as an absolute $p_T$ or as a fraction of the reconstructed lepton $p_T$
is available. For finer control, the thresholds that individual tracks or calorimeter cells are included 
in the isolation calculations can also be stated.

Whilst the above options allow reproduction of the vast majority of isolation variables used by the experiments
we also remind the user that both the tracking and calorimeter information is available if more control is required.

\subsubsection{Final state projections}

In order to simplify the basic acceptance cuts required on final state objects, \Checkmate\ 
employs a projection based system that is very similar to the one pioneered by Rivet \cite{Buckley:2010ar}. The idea 
is that a vector of final state objects is given to a function along with the the angular acceptance and minimum $p_T$.
A vector of the objects that passed the cuts is returned and this avoids multiple loops over final
state objects being needed in the analysis code. In addition, ATLAS often performs a cut on final state objects 
that fall into the transition region between the central and forward detectors and an optional projection
for this area is included.

\subsubsection{Overlap removal}

Another commonly required routine for LHC analyses is the removal of overlapping final state
objects. For example, since electrons deposit energy in the calorimeter they are also included
in the jet algorithm and must be removed from the list of jets in order not to double count momentum. In addition
many analyses also require an overlap removal between muons and jets as well. \Checkmate{} allows easy overlap removal 
where the user only has to state the cone size required for the operation, the object to be removed and the object to be 
kept.

\subsection{Kinematical Variables}
\label{sec:kinvar}

\Checkmate{} has access to both the Delphes kinematical variables and the \texttt{TLorentzVector} class from \Root. The Delphes\footnote{A full
list of the available functions is given at \url{https://cp3.irmp.ucl.ac.be/projects/delphes/wiki/WorkBook/RootTreeDescription}.} library
contains many of the most used kinematical parameters, like transverse momentum, energy and pseudo-rapidity, as well as non-kinematical properties, for example electric charge. In
addition the \texttt{TLorentzVector}\footnote{A full
list of the available functions is given at \url{https://root.cern.ch/root/html/TLorentzVector.html}.} class contains the 
full list of Lorentz variables and the possibility to calculate values from two separate vectors, e.g.\ relative angles.

\Checkmate{} also contains a number of popular LHC kinematical variables which are either hard-coded or sourced 
from other libraries (Oxbridge Kinetics Library \cite{Lester:1999tx,Barr:2003rg,Cheng:2008hk} and 
MctLib \cite{Tovey:2008ui,Polesello:2009rn}). For a more detailed review of many of variables that have now been
developed for high energy physics please see \cite{Barr:2010zj}.

\begin{itemize}
 \item $M_T$ \cite{vanNeerven:1982mz,Arnison:1983rp,Banner:1983jy,Smith:1983aa,Barger:1987du}: \\
    The transverse mass, $M_T$, allows for the reconstruction of a particle that decays semi-invisibly via 
    an end-point in the kinematical distribution.  We define,
    \begin{eqnarray}
	M_T^2 = m_{\mathrm{invis}}^2 + 2\, p^T\slashed{p}^T (1-\cos\Delta \phi)\,, \label{eq:m_T} 
    \end{eqnarray}
    where $p^T$ is the magnitude of the transverse momentum of the visible final state particle which is assumed to be 
    massless. $m_{\rm invis}$ is the mass and $\slashed{p}^T$ is the magnitude of the missing transverse 
    energy of the invisible particle and 
    $\Delta \phi$ is the azimuthal angle between the visible final state particle and the missing energy vector.
    
    $M_T$ is most commonly used to identify and measure the leptonic decay modes of the
    $W$ boson, $W\to \ell \nu$ in which case both of the final state particles, lepton and neutrino, can be considered massless.
 \item $M_{T2}$ \cite{Barr:2003rg,Cheng:2008hk}: \\
    In the case that an event contains two or more missing particles, an end-point in $M_T$ to measure
    the decaying particle mass will no longer exist. This is true when we for example consider the production
    of two particles that both decay to a final state that contains an invisible particle. In addition, for new 
    physics searches we are also often faced with the possibility that the masses of the invisible particles are unknown. If however the two invisible particles have the same mass, the space of possible momentum configurations is reduced and in fact can be numerically tested.
    To account for these  scenarios, the $M_{T2}$ variable was defined \cite{Barr:2003rg}
    which minimises over all possible partitions ($\slashed{\mathbf{p}}^T_1$ and $\slashed{\mathbf{p}}^T_2$) 
    of the vector missing energy ($\slashed{\mathbf{p}}^T$) between the two decay chains, 
    \begin{eqnarray}
	M_{T2}( m_{\chi}) = \underset{ \slashed{\mathbf{p}}^T_1+\slashed{\mathbf{p}}^T_2 =\slashed{\mathbf{p}}^T}{\mathrm{min}} [\mathrm{max}\{ M_{T} ({\bf p}^T_1,\slashed{\mathbf{p}}^T_1; m_{\chi}),\, M_{T} ({\bf p}^T_2,\slashed{\mathbf{p}}^T_2; m_{\chi}) \} ]\,.\label{eq:m_T2} 
    \end{eqnarray}
    Here, $\mathbf{p}^T_{1(2)}$ is the vector transverse momentum of the visible final state 
    particles, assumed to be massless, and $m_{\chi}$ is the mass of the invisible particle, which may be unknown and set 
    to some test value.
    
 \item $M_{T2}^{b\ell}$ \cite{Bai:2012gs}: \\
    A popular variable to efficiently discriminate scalar top partners which decay into top quarks and
    an invisible particle from the SM top background at the LHC is $M_{T2}^{b\ell}$.
    This is a modification of standard $M_{T2}$ to an asymmetric case where a single lepton and two 
    $b$-jets are reconstructed. The aim for $M_{T2}^{b\ell}$ is to reconstruct the top-quark mass in di-leptonic top
    decays where one lepton fails to be reconstructed. Thus, the decay chain with a reconstructed lepton should evaluate 
    $M_T$ with a massless neutrino, $m_{\nu}$, while the other chain --- assumed to be missing a lepton --- should 
    reconstruct the $W$ mass, $m_W$,
    \begin{eqnarray}
	M_{T2}^{b\ell} = \underset{ \slashed{\mathbf{p}}^T_1+\slashed{\mathbf{p}}^T_2 =\slashed{\mathbf{p}}^T}{\mathrm{min}} [\mathrm{max}\{ M_{T} ({\bf p}^T_{b_1} + {\bf p}^T_{\ell},\slashed{\mathbf{p}}^T_1; m_{\nu}),\, M_{T} ({\bf p}^T_{b_2},\slashed{\mathbf{p}}^T_2; m_{W}) \} ]\,.\label{eq:m^bl_T2} 
    \end{eqnarray}
    Here, $\mathbf{p}^T_{b_{1(2)}}$ is the vector transverse momentum of the $b$-jets and $\mathbf{p}^T_{\ell}$ is the vector transverse momentum of the lepton. 
    Commonly, both combinations of the $b$-jet momentum with the lepton momentum are computed and the smallest resulting $M_{T2}^{b\ell}$ is
    used.
    
 \item $M_{CT}$  \cite{Tovey:2008ui}: \\
     The $M_T$ variable (Eq.~\ref{eq:m_T}) is invariant under Lorentz--boosts. More specifically, it is invariant under \emph{co-linear transverse boosts}, i.e. if both particles are boosted by the same magnitude in the same transverse direction. 

A related observable called $M_{CT}$,  
    \begin{eqnarray}
	M^2_{CT} = m_{1}^2 + m_{2}^2 + 2\, (E^T_1 E^T_2 + {\bf p}^T_1\cdot{\bf p}^T_2)\,, \label{eq:m_CT} 
    \end{eqnarray}
    can be defined, where the transverse energy is defined as, $E^T_i = \sqrt{m_i^2+|{\bf p}_i^T|^2}$. Contrarily to $M_T$, it is invariant under \emph{contra-linear transverse boosts}, i.e.\ under boosts with equal magnitude but opposite direction in the transverse plane of the two final state particles. This is useful at a hadron collider
     since in the absence of initial state radiation (ISR), the production of a pair of equal mass particles will have equal back to back 
     transverse boosts. Moreover, these boosts will be unknown if the parent particle decays to an invisible particle.
     
 \item $M_{CT}$ corrected \cite{Polesello:2009rn}: \\ 
    A drawback of $M_{CT}$ is that the variable is not invariant under transverse boosts of the laboratory centre-of-mass
    frame. Consequently, transverse boosts from ISR will cause an end-point measured from $M_{CT}$ 
    to be smeared. However, a procedure to correct for such boosts was detailed in \cite{Polesello:2009rn} and is 
    implemented via MctLib.
    
 \item $M_{CT_{\bot}}$ and $M_{CT_{\lVert}}$ \cite{Matchev:2009ad}: \\
    In event topologies that allow for the reliable determination of ISR, two one dimensional decompositions of $M_{CT}$ can be
    made. We first define the decomposition of the final state momentum vectors into the components, perpendicular and parallel, 
    to the ISR transverse vector, ${\bf U}^T$,
    \begin{eqnarray}
	{\bf p}_i^{T_{\lVert}} & = & \frac{1}{|{\bf U}^T|^2} ({\bf p}_i^T\cdot{\bf U}^T){\bf U}^T\,, \label{eq:pT_perp} \\
	{\bf p}_i^{T_{\bot}} & = & {\bf p}_i^{T} - {\bf p}_i^{T_{\lVert}} =  \frac{1}{|{\bf U}^T|^2} {\bf U}^T \times ({\bf p}_i^T\times{\bf U}^T)\,. \label{eq:pT_par}
    \end{eqnarray}
    We can now define the one dimensional analogues of $M_{CT}$,
    \begin{eqnarray}
	M_{CT_{\bot}} & = & m_{1}^2 + m_{2}^2 + 2\, (E^{T_{\bot}}_1 E^{T_{\bot}}_2 + {\bf p}^{T_{\bot}}_1\cdot{\bf p}^{T_{\bot}}_2)\,, \label{eq:m_CT_perp}   \\
	M_{CT_{\lVert}} & = & m_{1}^2 + m_{2}^2 + 2\, (E^{T_{\lVert}}_1 E^{T_{\lVert}}_2 + {\bf p}^{T_{\lVert}}_1\cdot{\bf p}^{T_{\lVert}}_2)\,, \label{eq:m_CT_par} 
    \end{eqnarray}
    where we also define the corresponding energy decompositions,
    \begin{eqnarray}
	E^{T_{\bot}}_i = \sqrt{m_i^2+|{\bf p}_i^{T_{\bot}}|^2},\;\;\;\;\; E^{T_{\lVert}}_i = \sqrt{m_i^2+|{\bf p}_i^{T_{\lVert}}|^2}\,.
    \end{eqnarray}
    The perpendicular variable, $M_{CT_{\bot}}$, is calculated using vectors perpendicular to the ISR 
    and thus is invariant under boosts in the direction of ISR. 
     
 \item $\alpha_T$ \cite{Randall:2008rw,Khachatryan:2011tk}: \\
    A substantial background to missing energy searches with jet final states are pure QCD events where one jet has a large 
    mismeasurement. The mismeasurement can lead to a significant missing energy signal that could fake new physics with genuine
    missing energy and the variable $\alpha_T$, was designed to guard against this. For a di-jet system the variable is defined as,
    \begin{eqnarray}
	\alpha_T = \frac{E^T_{j_2}}{M_{T_j}}\,, \label{eq:alphaT}
    \end{eqnarray}
    where $E^T_{j_2}$ is the less energetic of the two jets and $M_{Tj}$ the transverse mass of the di-jet system,
    \begin{align}
	M^2_{Tj} = \left(\sum_{i=1}^{2} E^{T}_{j_i}\right)^2 -\left(\sum_{i=1}^{2} p^{x}_{j_i}\right)^2 -\left(\sum_{i=1}^{2} p^{y}_{j_i}\right)^2\,. \label{eq:m_Tj}   \\
    \end{align}
    For a perfectly back to back di-jet system, $\alpha_T=0.5$, but if one of the jets suffers a mismeasurement 
    in energy the resulting $\alpha_T$ will be lower. In contrast, signal like events that do not have a back to back
    topology but genuine missing energy can have values of $\alpha_T>0.5$.
    
    The variable is also generalised to multi-jet topologies,
    \begin{align}
	\alpha_T = \frac{H_T - \Delta H_T}{2 \sqrt{H_T^2 - \slashed{H}_T^2}}\,, \label{eq:alphaT_gen}
    \end{align}
    where,
    \begin{align}
	H_T = \sum_{i=1}^{n_{jet}} E^{T}_{j_i}\,,    \label{eq:HT}  \\
	\slashed{H}_T = \left| -\sum_{i=1}^{n_{jet}} {\bf p}^{T}_{j_i}\right|\,. \label{eq:HT_miss}
    \end{align}
    $\Delta H_T$ is found by forming two pseudo jets from all the reconstructed jets in the event. The combination is done by
    performing a scalar sum of the jet $E_T$ and choosing the jet combination that minimises the absolute $E_T$ 
    difference ($\Delta H_T$) between the two. 
    
 \item Razor \cite{Rogan:2010kb,Chatrchyan:2011ek,Khachatryan:2015pwa}: \\
    Another approach that has been designed to focus on the pair production of new states, e.g.\ scalar quark partners, that each decay into a 
    visible and an invisible particle, e.g.\ a neutralino and a jet, has been called `razor'. Two dimensionful variables are defined,
    \begin{align}
	M_R & =  \sqrt{(|{\bf p}_1| + |{\bf p}_2|)^2 - (p^z_1 + p^z_2)^2}\,,    \label{eq:MR}  \\
	M^R_T & =  \sqrt{ \frac{\slashed{p}^T(p^T_1 + p^T_2) - \slashed{\bf p}^T \cdot ({\bf p}^T_1 + {\bf p}^T_2)}{2} }\,. \label{eq:MRT}
    \end{align}
    where $M_R$ is designed to peak at the mass scale of the new physics and $M^R_T$ measures the transverse momentum imbalance.
    Combining the two values in a ratio,
    \begin{align}
        R = \frac{M^R_T}{M_R}\,,      \label{eq:Raz}   
    \end{align}
    one finds a variable that is expected to peak at $R\sim 0.5$ for two body decays. In the Standard Model, both $M_R$ and $R$ are
    expected to fall smoothly and thus new physics can appear as a bump on a falling background.
    
    As for $\alpha_T$, the variable can be generalised to topologies that contain more than two visible final state particles by
    first performing a clustering. In this case, two `megajets' are formed by summing the four-momentum of all combinations of final state 
    jets and leptons. The combination that minimises the sum of the invariant masses of the two megajets is then chosen.
    
\end{itemize}

\subsection{Validation}
\label{sec:validation}

An extremely important step on the way to recasting an LHC analysis is validation in order to ensure that all kinematical 
cuts are implemented correctly. LHC analyses now routinely provide information for particular hypothetical signal models
concerning the acceptance\footnote{Depending on
the analysis, this information can also be in the form of cross sections or simply Monte-Carlo event counts.} as each individual 
cut is applied in order. This information is vitally important when validating an analysis since it allows the acceptance
of each individual cut to be checked. 

In order to make the validation procedure as simple as possible, \Checkmate\ includes a specific syntax for recording cutflows that
can then be easily checked against the experimental publication. The syntax simply replaces the call \verb@countSignalEvent(<SR>)@
with \verb@countCutFlowEvent(<CF>)@ and generates completely separate cutflow results that can be quickly modified and do not 
interfere with the defined signal regions.

\subsection{Evaluation statistics}
\label{sec:stats}

One of the most convenient features of \Checkmate\ is the evaluation of the number of signal events and the corresponding 
calculation of all of the statistics required to determine if a particular model is excluded at the LHC or not. In fact 
the easiest way of running \Checkmate\ simply leads to the result \texttt{excluded} or \texttt{allowed} depending on the number of 
signal events.  More results and statistical variables are  available for more advanced users.
All of the above procedures are automatically evaluated for newly defined analyses with minimal user input required. 

For the implementation of existing LHC analyses, the user simply has to provide the number of background events expected 
in a particular signal region, the associated error and the actual number of observed events. All the normal \Checkmate\
evaluation routines are then available automatically and work with both weighted and unweighted signal events.

In the case that the user is developing a new analysis, the Standard Model backgrounds must first be evaluated before
the expected reach at the LHC can be calculated. For this the user simply runs \Checkmate\ as normal on the various 
backgrounds that are expected to contribute. The number of events normalised to the cross section is then evaluated 
by \Checkmate\ process by process for each signal region and simply summed to give the total background. Once this procedure
is complete the analysis can then be used as normal.

In addition, since we inherit all libraries contained within \Root, any of the statistical functions and 
histogramming features are easily accessible for use within analyses.

\section{Practical Examples}
\label{sec:pracex}
Often it is easiest to do a task oneself after one has seen it in a practical example. This is why in this section we will 
discuss the implementation of three practical examples in detail.

\subsection{Example 1: Recasting a simple analysis step by step}
In this part we will add an (already existing) analysis to the \Checkmate\ framework from the very first steps on. We assume that CheckMATE is already installed in the directory \texttt{<CM>}.

\subsubsection{Running the Analysis Manager}
If \verb@<CM>/bin/AnalysisManager@ does not yet exist, we run \verb@make AnalysisManager@ within \verb@<CM>@ to create 
it. Starting the AnalysisManager will open the introduction header and a first prompt:
\begin{Verbatim}[frame=leftline]

   ____ _               _    __  __    _  _____ _____ 
  / ___| |__   ___  ___| | _|  \/  |  / \|_   _| ____|
 | |   | '_ \ / _ \/ __| |/ / |\/| | / _ \ | | |  _|  
 | |___| | | |  __/ (__|   <| |  | |/ ___ \| | | |___ 
  \____|_| |_|\___|\___|_|\_\_|  |_/_/   \_\_| |_____|
                                           
                    /\  _  _ |   _. _  |\/| _  _  _  _  _ _ 
                   /--\| )(_||\/_)|_)  |  |(_|| )(_|(_)(-|  
                              /                     _/      
                                         
  What would you like to do? 
    -(l)ist all analyses,
    -(a)dd a new analysis to CheckMATE,
    -(e)dit analysis information,
    -(r)emove an analysis from CheckMATE
\end{Verbatim}
Apparently we want to add something new, so we hit \verb@a@:
\begin{Verbatim}[commandchars=\\\@\@,frame=leftline]
  This will collect all necessary information to create a full analysis and
  Takes care for the creation and implementation of the source files into the code.
  Please answer the following questions.
  Attention: Your input is NOT saved before you answered all questions!
\end{Verbatim}
The first block of questions gathers some general information on the analysis. At the beginning, you are asked for your name 
and your email address. These are printed on the very first lines of the analysis code to make sure that the right author 
can be contacted in case something goes wrong.
\begin{Verbatim}[commandchars=\\\!\!,frame=leftline]
  1. General Information to build analysis
    Your Name (to declare the analysis author): 
     \userinputcolor Guybrush Threepwood
    Your Email: 
     \userinputcolor threepwood@pirates.arr
\end{Verbatim}
Afterwards, some simple information on the analysis has to be provided. The name should be short but clear\footnote{Note that we only 
added the letter \texttt{X} at the end of the analysis name to prevent the AnalysisManager 
from overwriting the already implemented \texttt{atlas\_conf\_2013\_047} analysis data.} and should not contain any spaces, since this string defines the names for all analysis-specific files.
\begin{Verbatim}[commandchars=\\\@\@,frame=leftline]
    Analysis Name: 
     \userinputcolor atlas_conf_2013_047X 
\end{Verbatim}
The one-line description is the one used when someone types \verb@l@ in the AnalysisManager. It can also be found in the file \verb@list_of_analyses.dat@. The multiline description is printed on the top of every output file the respective analysis code creates:
\begin{Verbatim}[commandchars=\\\@\@,frame=leftline]
    Description (short, one line): 
     \userinputcolor ATLAS, 0 leptons + 2-6 jets + etmiss
    Description (long, multiple lines, finish with empty line:
     \userinputcolor ATLAS
     \userinputcolor ATLAS-CONF-2013-047
     \userinputcolor 0 leptons, 2-6 jets, etmiss
     \userinputcolor sqrt(s) = 8 TeV
     \userinputcolor int(L) = 20.3 fb^-1

\end{Verbatim}
The luminosity is important for the correct normalisation of the analysis results:
\begin{Verbatim}[commandchars=\\\@\@,frame=leftline]
    Luminosity (in fb^-1): 
     \userinputcolor 20.3
\end{Verbatim}
At the end of this block, the AnalysisManager will ask you whether you want to provide control regions for the given analysis. If we choose ``yes'', it will produce a second analysis source file which can be run separately from within \Checkmate. In our case, we are only interested in implementing signal regions and cutflows for which we do not need a second analysis file.
\begin{Verbatim}[commandchars=\\\@\@,frame=leftline]
    Do you plan to implement control regions to that analysis? [(y)es, (n)o)
     \userinputcolor n
\end{Verbatim}
Now we go more into the details of the analysis. As the next step, we have to list all signal regions the analysis 
provides. If we check out the actual analysis publication~\cite{ATLAS-CONF-2013-047}, we find five main signal 
regions A--E which sometimes are split into subregions L(oose), M(edium) or T(ight). We enter them one by one
\begin{Verbatim}[commandchars=\\\@\@,frame=leftline]
  2. Information on Signal Regions
    List all signal regions (one per line, finish with an empty line):
     \userinputcolor AL
     \userinputcolor AM
     \userinputcolor BM
     \userinputcolor BT
     \userinputcolor CM
     \userinputcolor CT
     \userinputcolor D
     \userinputcolor EL
     \userinputcolor EM
     \userinputcolor ET
     
\end{Verbatim}
In our case, we implement an experimental study for which the experimental observation and 
Standard Model expectation are known. If we want to create a new analysis, see later sections, we 
will not know these and have to find and enter them later.
\begin{Verbatim}[commandchars=\\\@\@,frame=leftline]
    Is the SM expectation B known? [(y)es, (n)o]?
       \userinputcolor y
\end{Verbatim}
Now things become a little more involved: for each of the above signal regions, we have to provide a set of numbers 
to which \Checkmate{} can compare the model prediction. In its most simple form, this includes the observed number 
of events \verb@obs@ and the Standard Model expectation \verb@bkg@ including error \verb@bkg_err@. Sometimes, the 
background error is split up into a statistical and systematical error and sometimes  the systematical error is given 
in asymmetrical $\pm$ form. The different input options also allow for 
the data to be given in all these different formats. In our 
example, Table 4 in ref.~\cite{ATLAS-CONF-2013-047} gives us the minimal set: \verb@obs@, \verb@bkg@ and \verb@bkg_err@. 
\begin{Verbatim}[commandchars=\\\@\@,frame=leftline]
    Information: We are now going to ask you which numbers you want to provide for each signal region. 
    The following items are possible:
        obs:                    Observed number of events
        bkg:                    Expected number of background events
        bkg_err:            Expected total error on bkg 
        bkg_errp:         Expected total upper error (in case of asymmetric errors)
        bkg_errm:         Expected total lower error (in case of asymmetric errors)
        bkg_err_stat: Expected statistical error on bkg
        bkg_err_sys:    Expected systematical error on bkg (in case of symmetric errors)
        bkg_errp_sys: Expected systematical upper error (in case of asymmetric errors)
        bkg_errm_sys: Expected systematical lower error (in case of asymmetric errors)
    Note that not all of these numbers have to be given (e.g. you don't have to give the total error if you give the individual stat and
    sys contributions)
    However, there are some requirements, about which you will be warned if you don't meet them (e.g. giving xyz_errp without xyz_errm)
    The standard, minimum set of information consists of obs, bkg and bkg_err
        List all categories you want to supply (one per line)
           \userinputcolor obs
           \userinputcolor bkg
           \userinputcolor bkg_err

    The set of information you entered is valid.
\end{Verbatim}
The AnalysisManager would reject the input list if it was not complete: for example, if we entered only the statistical error \verb@bkg_stat@ but no systematical error. 

Next, we have to enter the actual numbers for all the above categories for each and every one of the listed signal regions:
\begin{Verbatim}[commandchars=\\\@\@,frame=leftline]
        You now have to add the numbers for each of the given signal regions.
        Note that while you enter more numbers, the corresponding model independent
         95\% confidence limits for the items you have already entered are calculated
         in the background. 
        AL
            obs: 
               \userinputcolor 5333
            bkg: 
               \userinputcolor 4700
            bkg_err: 
               \userinputcolor 500
            S95obs and S95exp values are calculated internally (progress: 0 / 2)
        AM
            obs: 
               \userinputcolor 135
            bkg: 
               \userinputcolor 122
            bkg_err: 
               \userinputcolor 18
            S95obs and S95exp values are calculated internally (progress: 0 / 4)
        [...]
        ET
            obs: 
               \userinputcolor 5
            bkg: 
               \userinputcolor 2.9
            bkg_err: 
               \userinputcolor 1.8
            S95obs and S95exp values are calculated internally (progress: 3 / 20)
\end{Verbatim}
To allow for a fast statistical evaluation of the result in \Checkmate, it is convenient to 
first translate observation and expectation into a model independent upper limit on any new 
signal prediction \verb@S95@. These numbers are calculated internally by the AnalysisManager. As this 
calculation takes some time, it is queued and calculated in the background while the user 
enters the detector related information.

The next set of questions are all about detector level objects:
\begin{Verbatim}[commandchars=\\\@\@,frame=leftline]
    3. Settings for Detector Simulation
        3.1: Miscellaneous
            To which experiment does the analysis correspond? [(A)TLAS, (C)MS]
               \userinputcolor A
\end{Verbatim}
As a next step we have to enter information regarding lepton isolation criteria. Normally, these are defined within the experimental publications and are needed to distinguish signal leptons from leptons within jets. In our example, however, there are no signal leptons and therefore we do not require any particular isolation criteria for neither electrons, muons nor photons.
\begin{Verbatim}[commandchars=\\\@\@,frame=leftline]
        3.2: Electron Isolation
            Do you need any particular isolation criterion? [(y)es, (n)o]
               \userinputcolor n
        3.3: Muon Isolation
            Do you need any particular isolation criterion? [(y)es, (n)o]
               \userinputcolor n
        3.4: Photon Isolation
            Do you need any particular isolation criterion? [(y)es, (n)o]
               \userinputcolor n
\end{Verbatim}
Note that even though we entered \verb@no@, internally there will always be a soft isolation 
condition that cannot be overwritten by the user and which is automatically applied on electrons, muons 
and photons. This ensures that objects that would not have been reconstructed due to overlapping detector
 activity are automatically removed internally.

We now continue with the definition of jets. Reference~\cite{ATLAS-CONF-2013-047} tells us the exact properties we have to enter here:
\begin{Verbatim}[commandchars=\\\@\@,frame=leftline]
        3.5: Jets
            Which dR cone radius do you want to use for the FastJet algorithm?
               \userinputcolor 0.4
            What is the minimum pt of a jet? [in GeV]
               \userinputcolor 20
            Do you need a separate, extra type of jet? [(y)es, (n)o]
               \userinputcolor n
\end{Verbatim} 
The extra type of jet is needed in the rare case that a single analysis requires two different jet cone 
size, $\Delta R$, however it is not required in our example. 
The final questions relate to potential flavour tags we have to apply on jets. The given analysis 
considers $b$-tags with a reconstruction efficiency of 70\%.  We therefore simply enter
\begin{Verbatim}[commandchars=\\\@\@,frame=leftline]
            Do you want to use b-tagging? [(y)es, (n)o]
               \userinputcolor y
            b-Tagging 1:
                What is the signal efficiency to tag a b-jet? [in %]
                   \userinputcolor 70
            Do you need more b tags? [(y)es, (n)o]
               \userinputcolor n
            Do you want to use tau-tagging? [(y)es, (n)o]
               \userinputcolor n
\end{Verbatim}
Depending on how quickly we entered the above information, the AnalysisManger either finished the internally 
started S95 calculations or will wait until these are complete. When the evaluation ends, the results are shown and should be briefly checked to make sure 
that all numbers are sensible.
\begin{Verbatim}[commandchars=\\\@\@,frame=leftline]
    All necessary information has been entered. Before the AnalysisManager
     can create all required files, the internal S95obs and S95exp
     calculations have to finish. The calculation should take 10s up to a minute
     per point.
            ... done!
    Please check the below results for sanity. If anything looks
     suspicious, please contact the CheckMATE authors.
        obs    bkg    bkgerr    S95obs    S95exp
        5333    4700    500    1400    985
        135    122    18    50    40
        29    33    7    14    17
        4    2.4    1.4    6.9    5.8
        228    210    40    87    79
        0    1.6    1.4    3.0    3.4
        18    15    5    15    13
        166    113    21    90    54
        41    30    8    28    21
        5    2.9    1.8    8.2    6.6
        (Press any key to continue)
\end{Verbatim}
Note that due to numerical effects, your results might differ from the numbers above. Comparing the calculated numbers to the ones in Ref.~\cite{ATLAS-CONF-2013-047} tells us that the numbers are acceptable. Usually, the experimental numbers differ from the ones calculated by the AnalysisManager. This is mainly caused by different parametrisations of the background uncertainties, taking information on individual error sources into account to which \Checkmate\ does not have access. \Checkmate\ still prefers to calculate and use its own numbers as the exact same statistical routines for S95 are used for the proper CLs calculation in the \Checkmate\ evaluation routines such that the results of the two approaches are more consistent with each other.

We therefore accept the numbers and finish the AnalysisManager section:
\begin{Verbatim}[commandchars=\\\@\@,frame=leftline]
     - Variable values saved in /hdd/sandbox/managertest/data/atlas_conf_2013_047X_var.j
     - Created source file    /hdd/sandbox/managertest/tools/analysis/src/atlas_conf_2013_047X.cc
     - Created header file    /hdd/sandbox/managertest/tools/analysis/include/atlas_conf_2013_047X.h
     - Updated Makefile
     - Updated main source    main.cc
     - Reference file created
     - List of analyses updated
Analysis atlas_conf_2013_047X has been added successfully!
Run 'autoreconf; ./configure {parameters}; make' to compile the new sources.
You can find the list '{parameters}' you configured CheckMATE originally with at the beginning of the file 'config.log'.
\end{Verbatim}
\subsubsection{Looking at the Skeleton Code}
The AnalysisManager is mainly responsible to gather all the information for the detector simulation and the 
statistical evaluation sections of \Checkmate. What is still left to do is to define the analysis code that is used to analyse the events the user provides. In this section we describe how to do this by continuing our example of \verb@atlas_conf_2013_047X@. Note that the code we develop here differs slightly from the actual code implemented in \Checkmate\ since we, for the sake of understanding, change the order of some steps here and shorten the code where possible. It still gives the exact same results for the signal region tests though.

As the AnalysisManager has told us, source and header file for our analysis have already been created. These 
are already filled with skeleton code that properly compiles and embeds the analysis into the 
existing framework. What should concern us most is the source file which contains the actual 
analysis code. Every \Checkmate\ analysis contains three main functions.\footnote{The three-function structure is similar 
to the one used in the Rivet framework.}

\texttt{initialize()}: This function is called once at the beginning of an analysis. It is therefore used to setup overall information, initialise  variables and open files that are used throughout the analysis. By default, AnalysisManager 
data that is important for the analysis is already embedded:\footnote{Note that the actual source file includes 
many extra lines of explanatory comments which we have removed for this tutorial.}
\begin{Verbatim}[frame=leftline]
#include "atlas_conf_2013_047X.h"
// AUTHOR: Guybrush Threepwood
//  EMAIL: threepwood@pirates.arr
void Atlas_conf_2013_047x::initialize() {
  setAnalysisName("atlas_conf_2013_047X");          
  setInformation(""
    "# ATLAS\n"
         "# ATLAS-CONF-2013-047\n"
         "# 0 leptons, 2-6 jets, etmiss\n"
         "# sqrt(s) = 8 TeV\n"
         "# int(L) = 20.3 fb^-1\n"
  "");
  setLuminosity(20.3*units::INVFB);      
  ignore("towers"); 
  ignore("tracks"); 
  bookSignalRegions("AL;AM;BM;BT;CM;CT;D;EL;EM;ET;");
}
\end{Verbatim}
The \verb@setAnalysisName@ and \verb@setInformation@ functions define human readable headers and 
the prefix of all analysis-related output files. \verb@setLuminosity@ is used to properly normalise the input to a 
physical number of events given the cross section the user provides in the \Checkmate{} input card. This number 
can be used to rescale the overall output, e.g.\ to take a global event-cleaning efficiency into account.
\verb@ignore@ functions are for pure computing time optimisation reasons: they ignore information in the detector simulation 
to be read out if it is not required by the user. This is usually the case for calorimeter towers and tracker information which forms large datasets but which are rarely needed for the actual analysis.
Finally \verb@book@ functions are defined for \verb@SignalRegions@ and \verb@CutflowRegions@. These functions make sure that the respective \verb@_signal.dat@ and \verb@_cutflow.dat@ analysis output files contains numbers for each of the booked signal/cutflow regions. No matter in which order the signal regions are booked they will always be sorted alphabetically in the output.

\texttt{analyze()}: This function is the heart of any analysis: It is called once per event and contains the physics that is used to quantify the given data. Except for some comments and tips, this function does not contain any code yet and we will fill it in the next part of this tutorial.

\texttt{finalize()}: As the name suggests, this function is called once at the end of a given analysis. Usually this function is empty, however it can be used to e.g.\ free pointers that have been defined during \verb@initialize@ or to close some extra files that have been opened and filled before. 

Now that we have understood the structure let us start to implement the analysis code. 

\subsubsection{Selection of Final State Objects}
As the first step, we should define all the final state objects that we need 
and apply the appropriate kinematical cuts. The 
given analysis requires the following objects:
\begin{itemize}
\item missing energy
\item electrons with $\pT > 10$ GeV and $|\eta| < 2.47$ that pass minimal selection criteria
\item muons with $\pT > 10$ GeV and $|\eta| < 2.4$ reconstructed with combined data from tracker and muon spectrometer
\item photons with $\pT > 130$ GeV and $|\eta| < 2.47$ excluding $1.37<|\eta|<1.52$ and
\item jets with $\pT > 20$ GeV and $|\eta| < 2.8$.  
\end{itemize}
In the code, all the above objects are already predefined and we just have to select the right 
kinematic range using the \verb@filterPhaseSpace@ function of the analysis base class:
\begin{Verbatim}[commandchars=\\\@\@,frame=leftline]
void Atlas_conf_2013_047::analyze() {
  missingET->addMuons(muonsCombined);
 \userinputcolor electronsLoose = filterPhaseSpace(electronsLoose, 10., -2.47, 2.47);
 \userinputcolor muonsCombined = filterPhaseSpace(muonsCombined, 10., -2.4, 2.4);
 \userinputcolor jets = filterPhaseSpace(jets, 20., -2.8, 2.8);
 \userinputcolor photons = filterPhaseSpace(photons, 130., -2.47, 2.47, true);
\end{Verbatim}
The missing energy vector does not a priori contain muon contributions as 
different analyses can have different definitions of the construction of the missing momentum vector. We 
therefore add the contribution of all reconstructed muons by using the \verb@addMuons@ function. Furthermore, 
the area $1.37 \leq |\eta| \leq 1.52$ is often excluded for objects reconstructed in the electromagnetic
calorimeter. The \verb@filterPhaseSpace@ function therefore has an optional last parameter which --- if set to \verb@true@ --- ignores 
objects in exactly that region. 

These objects are then tested against different overlap conditions. In our case, these are defined as follows:
\begin{enumerate}
\item First, any jet with $\Delta R(jet,e) < 0.2$ to a nearby electron is removed.
\item Then, any electron and any muon with $\Delta R(\ell,jet) < 0.4$ is removed.
\end{enumerate}
Functions that take care of these removals are fortunately already available in the \verb@AnalysisBase@ class:
\begin{Verbatim}[commandchars=\\\@\@,frame=leftline]
 \userinputcolor jets = overlapRemoval(jets, electronsLoose, 0.2);
 \userinputcolor electronsLoose = overlapRemoval(electronsLoose, jets, 0.4);
 \userinputcolor muonsCombined = overlapRemoval(muonsCombined, jets, 0.4);
\end{Verbatim}
Beware that the order of these steps is crucial. Due to the setup of the detector simulation most electrons will also be reconstructed as jets. An overlap removal with respect to these two objects is therefore always necessary to avoid severe double counting issues.

\subsubsection{Event Selection}

Now that we have finished the object reconstruction step, we can start checking whether a given 
event fulfils the criteria of the given analysis. In our example, we 
should ignore events with leptons and hard photons. This is done very easily by using the \verb@return@ statement which ends the current run of the \verb@analyze()@ function and hence effectively vetoes the currently processed event:
\begin{Verbatim}[commandchars=\\\@\@,frame=leftline]
 \userinputcolor if(!photons.empty() || !electronsLoose.empty() || !muonsCombined.empty()) 
   \userinputcolor return;
\end{Verbatim}
We now start to look at the signal region criteria which are summarised in Table 1 in \cite{ATLAS-CONF-2013-047}. Firstly all 
signal regions require \etmiss$\geq 160$ GeV and at least two reconstructed jets. From these the leading 
jet must have a $\pT$ of least 130 GeV and the sub-leading jet $\pT>60$ GeV. All objects in \Checkmate{} analyses have a \verb@PT@ member that can be directly accessed for this purpose:
\begin{Verbatim}[commandchars=\\\@\@,frame=leftline]
\userinputcolor   if(missingET->PT < 160.0) 
\userinputcolor     return;   
\userinputcolor   if(jets.size() < 2 || jets[0]->PT < 130 || jets[1]->PT < 60)
\userinputcolor     return;
\end{Verbatim}
Note that all object vectors are automatically sorted with respect to $\pT$ such that leading and sub-leading jet are simply the 
first and the second object in the \verb@jet@ vector.\footnote{Standard C++ vectors do not catch out-of-bound 
indices. Users therefore have to ensure semantically that \texttt{jets[i]} is only accessed if \texttt{jets} contains at least $i+1$ members.}

Furthermore the angular separation of the missing momentum vector and the leading two jets must be at least 0.4. We make 
use of the \verb@P4()@ function of each object which returns the corresponding \verb@TLorentzVector@ object defined in \verb@ROOT@. This class comes with a large 
list of procedures to calculate 4-momentum parameters like the $\Delta \phi$ separation we need:
\begin{Verbatim}[commandchars=\\\@\@,frame=leftline]
\userinputcolor    if (fabs(jets[0]->P4().DeltaPhi(missingET->P4())) < 0.4)
\userinputcolor       return;
\userinputcolor    if( fabs(jets[1]->P4().DeltaPhi(missingET->P4())) < 0.4)
\userinputcolor       return;
\end{Verbatim}
There are more constraints on further `hard jets' with $\pT > 40$ GeV: If a third `hard jet' exists, it 
has to pass the above criterion too. For some signal regions, an additional constraint is applied if extra jets appear in the event. Here the looser constraint
that  $\Delta \phi > 0.2$ is applied to the fourth jet onwards:
\begin{Verbatim}[commandchars=\\\@\@,frame=leftline]
\userinputcolor    std::vector<Jet*> hardjets = filterPhaseSpace(jets, 40., -2.8, 2.8);
\userinputcolor    bool validThirdJet = (hardjets.size() < 3 || fabs(hardjets[2]->P4().DeltaPhi(missingET->P4())) > 0.4);
\userinputcolor    bool validMultiJet = validThirdJet;
\userinputcolor    for (int j = 3; j < hardjets.size(); j++) {
\userinputcolor        if (fabs(jets[j]->P4().DeltaPhi(missingET->P4())) < 0.2)
\userinputcolor            validMultiJet = false;
\userinputcolor    }
\end{Verbatim}
For our selection we need the ratio of \etmiss{} to the total effective mass $m_\text{eff}(Nj)$ as well as to $\sqrt{H_T}$. $H_T$ is defined to be the scalar $\pT$ sum of of all jets with $\pT > 40$ GeV whereas $m_\text{eff}(Nj)$ is the scalar $\pT$ sum of the leading $N$ jets plus \etmiss. There is also a requirement on $m_\text{eff}(\mathrm{incl}) := H_T + $\etmiss. We have to calculate these 
explicitly for all allowed number of jets.\footnote{The authors are aware that the conditions could be formulated 
more concisely by using the ternary \texttt{?} 
operator, e.g. \\ \texttt{double mEff4 = jets.size() >= 4 ? mEff3 + jets[3]->PT : 0;} For this manual we 
however preferred to use the easier to read version using \texttt{if} that most users should be familiar with.}
\begin{Verbatim}[commandchars=\\\@\@,frame=leftline]
\userinputcolor    double HT = 0.;
\userinputcolor    for(int j = 0; j < hardjets.size(); j++)
\userinputcolor       HT += hardjets[j]->PT;
\userinputcolor    double mEffincl = HT + missingET->PT;

\userinputcolor    double mEff2 = missingET->PT + jets[0]->PT + jets[1]->PT;
\userinputcolor    double rEff2 = missingET->PT/mEff2;

\userinputcolor    double mEff3 = 0;
\userinputcolor    if (jets.size() >= 3)
\userinputcolor        mEff3 =  mEff2 + jets[2]->PT;
\userinputcolor    double rEff3 = 0;
\userinputcolor    if (jets.size() >= 3)
\userinputcolor        rEff3 = missingET->PT/mEff3;

\userinputcolor    double mEff4 = 0;
\userinputcolor    if (jets.size() >= 4)
\userinputcolor        mEff4 =  mEff3 + jets[3]->PT;
\userinputcolor    double rEff4 = 0;
\userinputcolor    if (jets.size() >= 4)
\userinputcolor        rEff4 = missingET->PT/mEff4;
\userinputcolor    [...]

\userinputcolor    double rEffHT = missingET->PT/sqrt(HT);
\end{Verbatim}
The last requirement we have to check is that the momentum of the leading $N$ jets in a given signal region is larger than 60 GeV. For that we simply count the number of jets after cutting to the signal phase space:
\begin{Verbatim}[commandchars=\\\@\@,frame=leftline]
\userinputcolor    int nSignalJets = filterPhaseSpace(jets, 60., -2.8, 2.8).size();
\end{Verbatim}
With all these numbers at hand we can go ahead and check the various signal regions. Whenever an event passes the respective criteria, we just have to call the \verb@countSignalEvent@ function:
\begin{Verbatim}[commandchars=\\\@\@,frame=leftline]
\userinputcolor     if (validThirdJet) {
\userinputcolor         if(nSignalJets >= 2 && rEff2 > 0.2 && mEffincl > 1000.)
\userinputcolor             countSignalEvent("AL");
\userinputcolor         if(nSignalJets >= 2 && rEffHT > 15. && mEffincl > 1600.)
\userinputcolor             countSignalEvent("AM");
\userinputcolor         if(nSignalJets >= 3 && rEff3 > 0.3 && mEffincl > 1800.)
\userinputcolor             countSignalEvent("BM");
\userinputcolor         if(nSignalJets >= 3 && rEff3 > 0.4 && mEffincl > 2200.)
\userinputcolor             countSignalEvent("BT");
\userinputcolor     }
\userinputcolor     if (validMultiJet) {
\userinputcolor         if(nSignalJets >= 4 && rEff4 > 0.25 && mEffincl > 1200.)
\userinputcolor             countSignalEvent("CM");
\userinputcolor         if(nSignalJets >= 4 && rEff4 > 0.25 && mEffincl > 2200.)
\userinputcolor             countSignalEvent("CT");
\userinputcolor         if(nSignalJets >= 5 && rEff5 > 0.2 && mEffincl > 1600.)
\userinputcolor             countSignalEvent("D");
\userinputcolor         if(nSignalJets >= 6 && rEff6 > 0.15 && mEffincl > 1000.)
\userinputcolor             countSignalEvent("EL");
\userinputcolor         if(nSignalJets >= 6 && rEff6 > 0.2 && mEffincl > 1200.)
\userinputcolor             countSignalEvent("EM");
\userinputcolor         if(nSignalJets >= 6 && rEff6 > 0.25 && mEffincl > 1500.)
\userinputcolor             countSignalEvent("ET");
\userinputcolor     }
\end{Verbatim}
With these lines, we are actually done. The full code is listed 
in Fig.~\ref{fig:atlasconf2014047xfull}. 

\Checkmate\ then needs to be re-compiled with the following commands: \verb@autoreconf; ./configure {parameters};@ \verb@make@ within the \Checkmate\ main folder should compile the analysis and make it usable in the usual \Checkmate\ binary.

\begin{figure}
\begin{Verbatim}[commandchars=\\\@\@,frame=leftline,fontsize=\tiny]
void Atlas_conf_2013_047x::analyze() {
    missingET->addMuons(muonsCombined);
    electronsLoose = filterPhaseSpace(electronsLoose, 10., -2.47, 2.47);
    muonsCombined = filterPhaseSpace(muonsCombined, 10., -2.4, 2.4);
    jets = filterPhaseSpace(jets, 20., -2.8, 2.8);
    photons = filterPhaseSpace(photons, 130., -2.47, 2.47, true);

    jets = overlapRemoval(jets, electronsLoose, 0.2);
    electronsLoose = overlapRemoval(electronsLoose, jets, 0.4);
    muonsCombined = overlapRemoval(muonsCombined, jets, 0.4);

    if(!photons.empty() || !electronsLoose.empty() || !muonsCombined.empty())
        return;
    if(missingET->PT < 160.0)
        return;
    if(jets.size() < 2 || jets[0]->PT < 130 || jets[1]->PT < 60)
        return;
    if (fabs(jets[0]->P4().DeltaPhi(missingET->P4())) < 0.4)
        return;
    if( fabs(jets[1]->P4().DeltaPhi(missingET->P4())) < 0.4)
        return;

    std::vector<Jet*> hardjets = filterPhaseSpace(jets, 40., -2.8, 2.8);
    bool validThirdJet = (hardjets.size() < 3 || fabs(hardjets[2]->P4().DeltaPhi(missingET->P4())) > 0.4);
    bool validMultiJet = validThirdJet;
    for (int j = 3; j < hardjets.size(); j++) {
        if (fabs(hardjets[j]->P4().DeltaPhi(missingET->P4())) < 0.2)
           validMultiJet = false;
    }
    
    double HT = 0.;
    for(int j = 0; j < hardjets.size(); j++)
        HT += hardjets[j]->PT;
    double mEffincl = HT + missingET->PT;
    double mEff2 = missingET->PT + jets[0]->PT + jets[1]->PT;
    double rEff2 = missingET->PT/mEff2;

    double mEff3 = 0;
    if (jets.size() >= 3)
        mEff3 =  mEff2 + jets[2]->PT;
    double rEff3 = 0;
    if (jets.size() >= 3)
        rEff3 = missingET->PT/mEff3;

    double mEff4 = 0;
    if (jets.size() >= 4)
        mEff4 =  mEff3 + jets[3]->PT;
    double rEff4 = 0;
    if (jets.size() >= 4)
        rEff4 = missingET->PT/mEff4;

    double mEff5 = 0;
    if (jets.size() >= 5)
        mEff5 =  mEff4 + jets[4]->PT;
    double rEff5 = 0;
    if (jets.size() >= 5)
        rEff5 = missingET->PT/mEff5;

    double mEff6 = 0;
    if (jets.size() >= 6)
        mEff6 =  mEff5 + jets[5]->PT;
    double rEff6 = 0;
    if (jets.size() >= 6)
        rEff6 = missingET->PT/mEff6;
    double rEffHT = missingET->PT/sqrt(HT);

    int nSignalJets = filterPhaseSpace(jets, 60., -2.8, 2.8).size();    
    if (validThirdJet) {
        if(nSignalJets >= 2 && rEff2 > 0.2 && mEffincl > 1000.)
            countSignalEvent("AL");
        if(nSignalJets >= 2 && rEffHT > 15. && mEffincl > 1600.)
            countSignalEvent("AM");
        if(nSignalJets >= 3 && rEff3 > 0.3 && mEffincl > 1800.)
            countSignalEvent("BM");
        if(nSignalJets >= 3 && rEff3 > 0.4 && mEffincl > 2200.)
            countSignalEvent("BT");
    }
    if (validMultiJet) {
        if(nSignalJets >= 4 && rEff4 > 0.25 && mEffincl > 1200.)
            countSignalEvent("CM");
        if(nSignalJets >= 4 && rEff4 > 0.25 && mEffincl > 2200.)
            countSignalEvent("CT");
        if(nSignalJets >= 5 && rEff5 > 0.2 && mEffincl > 1600.)
            countSignalEvent("D");
        if(nSignalJets >= 6 && rEff6 > 0.15 && mEffincl > 1000.)
            countSignalEvent("EL");
        if(nSignalJets >= 6 && rEff6 > 0.2 && mEffincl > 1200.)
            countSignalEvent("EM");
        if(nSignalJets >= 6 && rEff6 > 0.25 && mEffincl > 1500.)
            countSignalEvent("ET");
    }
}
\end{Verbatim}
\caption{Full source code of the simplified reimplementation of \texttt{atlas\_conf\_2013\_047} into \Checkmate.}
\label{fig:atlasconf2014047xfull}
\end{figure} 
\pagebreak

\subsection{Example 2: Recasting a more complicated analysis}
The previous example was a rather simple search that did not contain many
complicated features within the actual analysis code. In this example 
we want to consider another analysis with 3 leptons and large missing transverse momentum in the final 
state, \texttt{atlas\_1402\_7029}, see Ref.~\cite{Aad:2014nua}. This analysis 
requires us to define specific isolation conditions for leptons, tag $b$ and $\tau$ jets and use 
advanced kinematic variables. We will also implement some cutflow steps 
to help track the signal efficiency when each cut is applied in turn.

\subsubsection{Detector Setting in AnalysisManager} 
We run the AnalysisManager as usual and add all the necessary information, including 
the data for all 24 signal regions (we do not go into more 
detail here: This step is identical to the analysis before). When it comes to the detector 
simulation part, things become more involved now. The given analysis requires two specific
electron isolation conditions: the first one needs the scalar sum of transverse momenta of tracks with $\pT > 0.4$ in the 
vicinity of $\Delta R \leq 0.3$ to the candidate to be at most 16 \% of the transverse momentum of the electron.\\
\begin{Verbatim}[commandchars=\\\@\@,frame=leftline]
    [...]
    3. Settings for Detector Simulation
        3.1: Miscellaneous
            To which experiment does the analysis correspond? [(A)TLAS, (C)MS]
               \userinputcolor A
        3.2: Electron Isolation
            Do you need any particular isolation criterion? [(y)es, (n)o]
               \userinputcolor y
            Isolation 1:
                Which objects should be considered for isolation? [(t)racks, (c)alo objects?]
                   \userinputcolor t
                What is the minimum pt of a surrounding object to be used for isolation? [in GeV]
                   \userinputcolor 0.4
                What is the dR used for isolation?
                   \userinputcolor 0.3
                Is there an absolute or a relative upper limit for the surrounding pt? [(a)bsolute, (r)elative]
                   \userinputcolor r
                What is the maximum pt ratio used for isolation?
                   \userinputcolor 0.16
\end{Verbatim}
A second condition checks the surrounding calorimeter cells in the same region to contain at 
most 18 \% of the candidate's momentum. The $p_T^{\mathrm{min}}$ of the considered cells is set to be $0.1$ so that random cell noise is ignored.
\begin{Verbatim}[commandchars=\\\@\@,frame=leftline]
            Do you need more isolation criteria? [(y)es, (n)o]
               \userinputcolor y
            Isolation 2:
                Which objects should be considered for isolation? [(t)racks, (c)alo objects?]
                   \userinputcolor c
                What is the minimum pt of a surrounding object to be used for isolation? [in GeV]
                   \userinputcolor 0.1
                What is the dR used for isolation?
                   \userinputcolor 0.3
                Is there an absolute or a relative upper limit for the surrounding pt? [(a)bsolute, (r)elative]
                   \userinputcolor r
                What is the maximum pt ratio used for isolation?
                   \userinputcolor 0.18
\end{Verbatim}
That is all for the electrons. For the muons we also have a condition which is very similar to the first electron condition but with different numbers:
\begin{Verbatim}[commandchars=\\\@\@,frame=leftline]
        3.3: Muon Isolation
            Do you need any particular isolation criterion? [(y)es, (n)o]
               \userinputcolor y
            Isolation 1:
                Which objects should be considered for isolation? [(t)racks, (c)alo objects?]
                   \userinputcolor t
                What is the minimum pt of a surrounding object to be used for isolation? [in GeV]
                   \userinputcolor 1.0
                What is the dR used for isolation?
                   \userinputcolor 0.3
                Is there an absolute or a relative upper limit for the surrounding pt? [(a)bsolute, (r)elative]
                   \userinputcolor r
                What is the maximum pt ratio used for isolation?
                   \userinputcolor 0.12
            
            Do you need more isolation criteria? [(y)es, (n)o]
               \userinputcolor n
\end{Verbatim}
Photons are not needed in our analysis:
\begin{Verbatim}[commandchars=\\\@\@,frame=leftline]
        3.4: Photon Isolation
            Do you need any particular isolation criterion? [(y)es, (n)o]
               \userinputcolor n
\end{Verbatim}
The analysis demands vetoes against jets that originated from $b$-quarks. For that, a $b$-tagging algorithm is 
used that is tuned to a working point signal efficiency of 80 \%. Furthermore, `medium' $\tau$-tags are used to 
identify signal events. We therefore have to include these objects in the AnalysisManager:
\begin{Verbatim}[commandchars=\\\@\@,frame=leftline]
        3.5: Jets
            Which dR cone radius do you want to use for the FastJet algorithm?
               \userinputcolor 0.4
            What is the minimum pt of a jet? [in GeV]
               \userinputcolor 20.0
            Do you need a separate, extra type of jet? [(y)es, (n)o]
               \userinputcolor n
            Do you want to use b-tagging? [(y)es, (n)o]
               \userinputcolor y
            b-Tagging 1:
                What is the signal efficiency to tag a b-jet? [in %]
                    \userinputcolor 80
            Do you need more b tags? [(y)es, (n)o]
                \userinputcolor n
            Do you want to use tau-tagging? [(y)es, (n)o]
                \userinputcolor y
\end{Verbatim}
Note that when $\tau$-tagging is activated, all three working points `loose', `medium' and `tight' are automatically tagged and available in the analyses. This step will finish the AnalysisManager part and create the analysis skeleton file as usual.

\subsubsection{Analysis: Setup}

With all the parameters set, we can continue with the analysis code.\footnote{It should be noted that the implementation 
of the analysis we describe here slightly differs from the public version used 
in \Checkmate{}, as the public version considers more cutflows which sometimes require a different ordering.} Since we also want to check the 
cutflow of our analysis, we should book cutflow regions to make 
sure the \verb@_cutflow.dat@ file is produced at the end of a \Checkmate\ analysis run. For this we need to examine
 the \verb@initialize()@ function in our analysis skeleton file. Here all 
the signal regions we defined above are already booked:
\begin{Verbatim}[commandchars=\\\@\@,frame=leftline]
  bookSignalRegions("SR0taua01;SR0taua02;SR0taua03;SR0taua04;SR0taua05;...
\end{Verbatim} 
For cutflows we add a similar line with all the desired cutflow region names. In this 
tutorial, let us check the number of events we lose after the trigger, after the event selection and how many we get with 0, 1 or 2 $\tau$'s:
\begin{Verbatim}[commandchars=\\\@\@,frame=leftline]
  \userinputcolor bookCutflowRegions("CR0_All;CR1_Trigger;CR2_Selection;CR3_0Tau;CR3_1Tau;CR3_2Tau;")
\end{Verbatim} 
Note that regions are automatically sorted alphabetically. We therefore constructed the names such that they will appear in logical order in the output file.
\subsubsection{Analysis: Event Cleaning}
As before, we focus on the \verb@analyze()@ function in the analysis source file. We start with the definition of the 
interesting final state objects and the overlap removals, whose proper definitions can be 
found in~\cite{Aad:2014nua}. Note that here we have to consider both `medium' and `tight' electrons, where the former are 
used for the event cleaning at the beginning and the latter are necessary for the signal region tests later. We also have 
to remove electrons that lie in the vicinity of each other, i.e.\ if two electrons are closer than $\Delta R = 0.1$, the 
 one with lower energy should be removed. This is done using the same \verb@overlapRemoval()@ function 
as before, but now with only one particle vector parameter given.\footnote{Note that \texttt{overlapRemoval(electronsMedium, electronsMedium, 0.1)} must 
not be used: It will always return an empty list as each electron in the first vector 
finds an `overlapping' $\Delta R = 0$ electron in the second vector, namely itself.}
\begin{Verbatim}[commandchars=\\\@\@,frame=leftline]
void Atlas_1402_7029::analyze() {
   missingET->addMuons(muonsCombined);  
  \userinputcolor electronsMedium = filterPhaseSpace(electronsMedium, 10., -2.47, 2.47, false);
  \userinputcolor electronsTight = filterPhaseSpace(electronsTight, 10., -2.47, 2.47, false);
  \userinputcolor muonsCombined = filterPhaseSpace(muonsCombined, 10., -2.4, 2.4);
  \userinputcolor jets = filterPhaseSpace(jets, 20., -2.5, 2.5);  

  \userinputcolor electronsMedium = overlapRemoval(electronsMedium, 0.1);
  \userinputcolor electronsTight = overlapRemoval(electronsTight, 0.1);
  \userinputcolor jets = overlapRemoval(jets, electronsMedium, 0.2);
  \userinputcolor electronsMedium = overlapRemoval(electronsMedium, jets, 0.4);
  \userinputcolor electronsTight = overlapRemoval(electronsTight, jets, 0.4);
  \userinputcolor muonsCombined = overlapRemoval(muonsCombined, jets, 0.4);
\end{Verbatim}
Next follows an analysis-specific `resonance removal' which forces leptons that form opposite-sign same-flavour pairs with an invariant mass smaller than 12~GeV to be removed. We can do this using a simple double-loop. For electrons, we only need to test signal electrons (`tight') but they must not form an invariant mass pair with `medium' electrons either. Since `tight' electrons are a subset of `medium' electrons this will also remove any pairs formed by two `tight' electrons.
\begin{Verbatim}[commandchars=\\\@\@,frame=leftline]
  \userinputcolor std::vector<Electron*> noResonanceElecs;
  \userinputcolor for (int t = 0; t < electronsTight.size(); t++) {
      \userinputcolor bool valid = true;
      \userinputcolor for (int m = 0; m < electronsMedium.size(); m++) {
          \userinputcolor if (electronsMedium[m]->Charge*electronsTight[t]->Charge > 0)
              \userinputcolor continue;  
          \userinputcolor if ( (electronsMedium[m]->P4() + electronsTight[t]->P4()).M() < 12.)
              \userinputcolor valid = false;
      \userinputcolor }
      \userinputcolor if (valid)
          \userinputcolor noResonanceElecs.push_back(electronsTight[t]);
  \userinputcolor }
\end{Verbatim}
We do the same for muons and store the cleaned vectors:
\begin{Verbatim}[commandchars=\\\@\@,frame=leftline]
  \userinputcolor std::vector<Muon*> noResonanceMuons;
  \userinputcolor for (int t = 0; t < muonsCombined.size(); t++) {
      \userinputcolor bool valid = true;
      \userinputcolor for (int m = 0; m < muonsCombined.size(); m++) {
          \userinputcolor if (muonsCombined[m]->Charge*muonsCombined[t]->Charge > 0)
              \userinputcolor continue; 
          \userinputcolor if ( (muonsCombined[m]->P4() + muonsCombined[t]->P4()).M() < 12.)
              \userinputcolor valid = false;
\userinputcolor       }
\userinputcolor       if (valid)
\userinputcolor           noResonanceMuons.push_back(muonsCombined[t]);
\userinputcolor   }
\userinputcolor   electronsTight = noResonanceElecs;
\userinputcolor   muonsCombined = noResonanceMuons;
\end{Verbatim}
We further select our signal final state objects by applying our defined isolation conditions. Note that applying the function without any arguments will apply all stored conditions on the respective object, i.e.\ both conditions for the electron and the single condition for the muon:
\begin{Verbatim}[commandchars=\\\@\@,frame=leftline]
\userinputcolor   electronsTight = filterIsolation(electronsTight);
\userinputcolor   muonsCombined = filterIsolation(muonsCombined);
\end{Verbatim}
Let us now keep track of all events we start with in the cutflow:
\begin{Verbatim}[commandchars=\\\@\@,frame=leftline]
\userinputcolor   countCutflowEvent("CR0_All");
\end{Verbatim}
\subsubsection{Analysis: Event Trigger}
The next step tests if the event could have passed at least one of the 
applied trigger conditions. For the present analysis, various triggers with different momentum thresholds for the 
considered objects are tested. We do not consider any turn-on curves but simply 
apply a 100 \% trigger efficiency if there are objects that pass the respective $p_T^{\mathrm{min}}$ criteria:
\begin{Verbatim}[commandchars=\\\@\@,frame=leftline]
\userinputcolor   bool trigger = false;
\userinputcolor   if( electronsTight.size() > 0 && electronsTight[0]->PT > 25.)
\userinputcolor     trigger = true;
\userinputcolor   else if( muonsCombined.size() > 0 && muonsCombined[0]->PT > 25.) 
\userinputcolor     trigger = true;
\userinputcolor   else if( electronsTight.size() > 1 && electronsTight[0]->PT > 14. && electronsTight[1]->PT > 14.)
\userinputcolor     trigger = true;
\userinputcolor   else if( electronsTight.size() > 1 && electronsTight[0]->PT > 25. && electronsTight[1]->PT > 10.)
\userinputcolor     trigger = true;
\userinputcolor   else if( muonsCombined.size() > 1 && muonsCombined[0]->PT > 14. && muonsCombined[1]->PT > 14.)
\userinputcolor     trigger = true;
\userinputcolor   else if( muonsCombined.size() > 1 && muonsCombined[0]->PT > 18. && muonsCombined[1]->PT > 10.)
\userinputcolor     trigger = true;
\userinputcolor   else if( electronsTight.size() > 0 && muonsCombined.size() > 0 && electronsTight[0]->PT > 18. && muonsCombined[0]->PT > 10.)
\userinputcolor     trigger = true;
\userinputcolor   else if( electronsTight.size() > 0 && muonsCombined.size() > 0 && electronsTight[0]->PT > 10. && muonsCombined[0]->PT > 18.)
\userinputcolor       trigger = true;
\userinputcolor   if( !trigger )
\userinputcolor       return;
\end{Verbatim}
Our cutflow should tell us how many events pass this stage:
\begin{Verbatim}[commandchars=\\\@\@,frame=leftline]
\userinputcolor   countCutflowEvent("CR1_Trigger");
\end{Verbatim}
\subsubsection{Analysis: Event Selection}
The present three lepton analysis considers signal regions with tau leptons and for that purpose we have to consider the jets that have been tagged as ``medium'' tau jets by using the \verb@checkTauTag@ function. It is important to note that tagged jets a priori do not have to pass conditions on the number and charge of reconstructed tracks. In the analysis we have to explicitly ensure this by testing the charge of the tested jets. Events then have to contain exactly three signal leptons ($e$, $\mu$, $\tau$) but not exactly three $\tau$ only:
\begin{Verbatim}[commandchars=\\\@\@,frame=leftline]
\userinputcolor   std::vector<Jet*> tauJets;
\userinputcolor   for(int j = 0; j < jets.size(); j++) {
\userinputcolor       if( checkTauTag(jets[j], "medium") && fabs(jets[j]->Charge) == 1) 
\userinputcolor           tauJets.push_back(jets[j]);      
\userinputcolor   }
\userinputcolor   if( ( electronsTight.size() + muonsCombined.size() + tauJets.size() ) != 3 )
\userinputcolor         return; 
\userinputcolor   if(  electronsTight.size() + muonsCombined.size() == 0 )
\userinputcolor       return;  
\end{Verbatim} 
Furthermore no $b$-jet should be present in the final state. We can test this by applying the \verb@checkBTag()@ function on a given jet candidate and, in our analysis, veto the event if any of these tests return true:
\begin{Verbatim}[commandchars=\\\@\@,frame=leftline]
\userinputcolor   for (int i = 0; i < jets.size(); i++) {
\userinputcolor       if( checkBTag(jets[i]))
\userinputcolor           return;
\end{Verbatim}
Again we want to know how many events pass all of the above cuts:
\begin{Verbatim}[commandchars=\\\@\@,frame=leftline]
\userinputcolor   countCutflowEvent("CR2_Selection");
\end{Verbatim}
\subsubsection{Analysis: Signal Region Categorisation}
We now start categorising the event into the different signal regions. From this point on we will often consider both 
electrons and muons on the same footing. In certain cases the same treatment will also be applied to taus. It would therefore be convenient if 
we combined all the three leptons into a common vector which we can use if the kinematical cut is flavour independent. Unfortunately, the three objects are described by different C++ classes and hence cannot be combined trivially. However,
within the analysis code there is a new class \verb@FinalStateObject@ into which electrons, muons, jets and even the missing 
momentum vector can be transformed. We therefore define a vector of \verb@FinalStateObject@ objects and choose to fill it 
with the three different kinds of lepton we have in the final state.
\begin{Verbatim}[commandchars=\\\@\@,frame=leftline]
\userinputcolor   std::vector<FinalStateObject*> leptons;
\userinputcolor   for(int e = 0; e < electronsTight.size(); e++) {
\userinputcolor      FinalStateObject* lep = newFinalStateObject(electronsTight[e]);
\userinputcolor      leptons.push_back(lep);
\userinputcolor   }  
\userinputcolor   for(int m = 0; m < muonsCombined.size(); m++) {
\userinputcolor      FinalStateObject* lep = newFinalStateObject(muonsCombined[m]);
\userinputcolor      leptons.push_back(lep);
\userinputcolor   }
\userinputcolor   for(int t = 0; t < tauJets.size(); t++) {
\userinputcolor      FinalStateObject* lep = newFinalStateObject(tauJets[t]);
\userinputcolor      leptons.push_back(lep);
\userinputcolor   }
\end{Verbatim}
Note that this ordering ensures that the tau objects always come last. With this vector we can easily test the further condition that all leptons have to be separated by at least $\Delta R \geq 0.3$, without distinguishing electrons, muons or tau--jets:
\begin{Verbatim}[commandchars=\\\@\@,frame=leftline]
\userinputcolor    if (leptons[0]->P4().DeltaR(leptons[1]->P4()) < 0.3)
\userinputcolor        return;
\userinputcolor    if (leptons[0]->P4().DeltaR(leptons[2]->P4()) < 0.3)
\userinputcolor        return;
\userinputcolor    if (leptons[1]->P4().DeltaR(leptons[2]->P4()) < 0.3)
\userinputcolor        return;
\end{Verbatim}
The main difference between the signal region definitions is the number of tau jets in the final state. We will use our cutflow regions to check how many events in general have 0, 1 or 2 taus. Let us start with the definition of the two signal regions involving two tau jets:
\begin{Verbatim}[commandchars=\\\@\@,frame=leftline]
\userinputcolor switch(tauJets.size()) {
\userinputcolor     case 2: {
\userinputcolor         countCutflowEvent("CR3_2tau");
\end{Verbatim}
The first one of these, \verb@SR2taua@, requires at least 50~GeV missing energy in the event. Furthermore, it 
tests the  \verb@mT2@ variable, see \ref{sec:kinvar} for more information.  This variable 
 requires two \verb@TLorentzVectors@ arguments, i.e.\ two four-momenta, 
and the mass of the invisible particle that is typically assumed to be 0, to reject SM background due to leptonic $W$ decays. Signal 
region \verb@SR2taua@ then requires the maximum \verb@mT2@ out of all combinations of the three signal leptons to be at 
least 100 GeV. With our above universal \verb@leptons@ vector, this is a very easy exercise:
\begin{Verbatim}[commandchars=\\\@\@,frame=leftline]
\userinputcolor         double mT2_1 = mT2(leptons[0]->P4(), leptons[1]->P4(), 0.);
\userinputcolor         double mT2_2 = mT2(leptons[0]->P4(), leptons[2]->P4(), 0.);
\userinputcolor         double mT2_3 = mT2(leptons[1]->P4(), leptons[2]->P4(), 0.);
\userinputcolor         double mT2max = std::max(std::max(mT2_1, mT2_2), mT2_3);
\userinputcolor         if (missingET->PT > 50. && mT2max > 100)
\userinputcolor             countSignalEvent("SR2taua");
\end{Verbatim}
The other signal region, {\verb@SR2taub@}, requires different total missing 
energy, an opposite sign tau pair with their scalar $\pT$ sum  to be larger than 110~GeV and their invariant mass 
to be between 70~GeV and 120~GeV:
\begin{Verbatim}[commandchars=\\\@\@,frame=leftline]
\userinputcolor         double mtautau = (tauJets[0]->P4()+tauJets[1]->P4()).M();
\userinputcolor         double sumpt = tauJets[0]->PT + tauJets[1]->PT;
\userinputcolor         if (missingET->PT > 60. && tauJets[0]->Charge*tauJets[1]->Charge < 0 && sumpt > 110. && mtautau > 70. && mtautau < 120.)
\userinputcolor             countSignalEvent("SR2taub");           
\end{Verbatim}
We continue with the $1\tau$ signal region:
\begin{Verbatim}[commandchars=\\\@\@,frame=leftline]
\userinputcolor         break;
\userinputcolor     }
\userinputcolor     case 1: {
\userinputcolor         countCutflowEvent("CR3_1tau");
\end{Verbatim}
Here we have to find the $\tau$ and light lepton invariant mass combination that lies closest 
to the Higgs boson mass and the invariant mass of the light lepton pair. Since the tau jet is always the third object in the \verb@leptons@ vector --- caused by the order we filled it above --- we know which combinations to test. The invariant mass is easily calculated using the \verb@TLorentzVector@ internal \verb@M()@ function:
\begin{Verbatim}[commandchars=\\\@\@,frame=leftline]
\userinputcolor         double mTL_1 = (leptons[0]->P4() + leptons[2]->P4()).M();
\userinputcolor         double mTL_2 = (leptons[1]->P4() + leptons[2]->P4()).M();
\userinputcolor         double mTL = fabs(mTL_1 - 125.) < fabs(mTL_2 - 125.) ? mTL_1 : mTL_2;
\userinputcolor         double mLL   = (leptons[0]->P4() + leptons[1]->P4()).M();
\end{Verbatim}
The light leptons must have same charge and the event is 
vetoed if an electron pair reconstruct an invariant 
mass compatible with the $Z$ boson:
\begin{Verbatim}[commandchars=\\\@\@,frame=leftline]
\userinputcolor         bool samesignl = (leptons[0]->Charge*leptons[1]->Charge > 0);
\userinputcolor         bool zElectronVeto = (leptons[0]->Type == "electron" && leptons[1]->Type == "electron" && mLL > 81.2 && mLL < 101.2);
\end{Verbatim}
Lastly, these combine with requirements on the missing transverse 
momentum and the transverse momenta of the light leptons:
\begin{Verbatim}[commandchars=\\\@\@,frame=leftline]
\userinputcolor         double sumpt = leptons[0]->PT + leptons[1]->PT;
\userinputcolor         if (missingET->PT > 50. && leptons[0]->PT > 30. && leptons[1]->PT > 30. && sumpt > 70. && mTL < 120. && samesignl && !zElectronVeto)
\userinputcolor           countSignalEvent("SR1tau"); 
\userinputcolor         break;
\userinputcolor     }
\userinputcolor     case 0: {
\userinputcolor         countCutflowEvent("CR3_0tau");
\end{Verbatim}
For this signal region, we first have to check if there is a same-flavour opposite-sign (SFOS) pair.
\begin{Verbatim}[commandchars=\\\@\@,frame=leftline]
\userinputcolor         bool sfos = false;
\userinputcolor         double msfos = -1E10;
\userinputcolor         if (leptons[0]->Charge * leptons[1]->Charge < 0 && leptons[0]->Type == leptons[1]->Type)
\userinputcolor            sfos = true;
\userinputcolor         else if (leptons[0]->Charge * leptons[2]->Charge < 0 && leptons[0]->Type == leptons[2]->Type)
\userinputcolor            sfos = true;
\userinputcolor         else if (leptons[2]->Charge * leptons[1]->Charge < 0 && leptons[2]->Type == leptons[1]->Type)
\userinputcolor            sfos = true;
\end{Verbatim}
Let us now continue with the test of \verb@SR0taub@ which requires no SFOS pair. In that case, we have to require 
that there is at least one lepton of different flavour (OF) and that the product of all charges is 
negative. We then have to find the OF combination which yields the smallest relative polar angle $\Delta \Phi$. Note that our 
construction has produced a \verb@leptons@ vector of either of the four possibilities $eee$, $ee\mu$, $e\mu\mu$ or $\mu\mu\mu$ so one does 
not have to test all combinations to find OF pairs:
\begin{Verbatim}[commandchars=\\\@\@,frame=leftline]
\userinputcolor         if (!sfos && leptons[0]->Type != leptons[2]->Type && fabs(leptons[0]->Charge + leptons[1]->Charge + leptons[2]->Charge) == 1) {
\userinputcolor            double deltaPhi1 = fabs(leptons[0]->P4().DeltaPhi(leptons[2]->P4()));
\userinputcolor            double deltaPhi2 = 0;
\userinputcolor            if (leptons[0]->Type != leptons[1]->Type)
\userinputcolor                deltaPhi2 = fabs(leptons[0]->P4().DeltaPhi(leptons[1]->P4()));
\userinputcolor            else             
\userinputcolor                deltaPhi2 = fabs(leptons[1]->P4().DeltaPhi(leptons[2]->P4()));
\userinputcolor            double mindeltaPhi = std::min(deltaPhi1, deltaPhi2);
\userinputcolor            if(missingET->PT > 50. && leptons[0]->PT > 20. && leptons[1]->PT > 20. && leptons[2]->PT > 20. && mindeltaPhi <= 1.0)
\userinputcolor                countSignalEvent("SR0taub"); 
\userinputcolor         }
\end{Verbatim}
For the SFOS signal region we have to find the SFOS combination with invariant mass closest to the Z boson. We need this invariant mass and the transverse mass of the remaining lepton:
\begin{Verbatim}[commandchars=\\\@\@,frame=leftline]
\userinputcolor         else if(sfos) {
\userinputcolor             double msfos = -1E10;
\userinputcolor             double mTthird = 0;
\userinputcolor             if (leptons[0]->Charge * leptons[1]->Charge < 0 && leptons[0]->Type == leptons[1]->Type) {
\userinputcolor                 double minv = (leptons[0]->P4() + leptons[1]->P4()).M()
\userinputcolor                 if (fabs(minv - 92.) < fabs(msfos - 92.)) {
\userinputcolor                    msfos = minv;
\userinputcolor                    mtThird = mT(leptons[2]->P4(), missingET->P4());
\userinputcolor                 }
\userinputcolor             }
\userinputcolor             if (leptons[0]->Charge * leptons[2]->Charge < 0 && leptons[0]->Type == leptons[2]->Type) {
\userinputcolor                 double minv = (leptons[0]->P4() + leptons[2]->P4()).M()
\userinputcolor                 if (fabs(minv - 92.) < fabs(msfos - 92.)) {
\userinputcolor                    msfos = minv;
\userinputcolor                    mtThird = mT(leptons[1]->P4(), missingET->P4());
\userinputcolor                 }
\userinputcolor             }
\userinputcolor             else if (leptons[2]->Charge * leptons[1]->Charge < 0 && leptons[2]->Type == leptons[1]->Type){
\userinputcolor                 double minv = (leptons[2]->P4() + leptons[1]->P4()).M()
\userinputcolor                 if (fabs(minv - 92.) < fabs(msfos - 92.)) {
\userinputcolor                     msfos = minv;
\userinputcolor                     mtThird = mT(leptons[0]->P4(), missingET->P4());
\userinputcolor                 }
\userinputcolor             }
\end{Verbatim}
Lastly, some signal regions specifically veto if the three lepton invariant mass lies within the Z boson mass window:
\begin{Verbatim}[commandchars=\\\@\@,frame=leftline]
\userinputcolor             double m3l = (leptons[0]->P4() + leptons[1]->P4() + leptons[2]->P4()).M();
\end{Verbatim}
The analysis ends with a binning of missing energy, \verb@msfos@ and \verb@mtThird@
\begin{Verbatim}[commandchars=\\\@\@,frame=leftline]
\userinputcolor      	    if (12. < msfos && msfos < 40) {
\userinputcolor      	        if (0. <= mtThird && mtThird < 80) {
\userinputcolor      	            if (50. <= missingET->PT && missingET->PT < 90.)
\userinputcolor      	                countSignalEvent("SR0taua01");     	              
\userinputcolor      	            else if (90. <= missingET->PT)
\userinputcolor      	                countSignalEvent("SR0taua02");
\userinputcolor      	        }
\userinputcolor      	        else if (80. <= mtThird ) {
\userinputcolor      	          ...
\end{Verbatim}
\subsubsection{Analysis Testing}
To check the implementation of our analysis, let us run \Checkmate{} on a simple model that is also 
analysed within \cite{Aad:2014nua}: a simplified supersymmetric scenario with direct production 
of the lightest chargino and the second lightest neutralino, both with mass $m_{\tilde{\chi}^{\pm}_1} = m_{\tilde{\chi}^0_2} = 625 $ GeV. These 
both decay into the lightest neutralino with mass $m_{\tilde{\chi}^0_1} = 375 $ GeV and two leptons via an intermediate 
slepton with mass $m_{\tilde{\ell}} = 500 $ GeV. By charge conservation this will lead to an odd number of charged 
leptons in the final state and is consequently a good candidate for our underlying three lepton study. We take the 
corresponding slha spectrum file \cite{Skands:2003cj,Allanach:2008qq} and use 
Prospino \cite{Beenakker:1999xh,Beenakker:1996ed} to calculate the 
production cross section to be $\sigma \approx 2.893 $ pb. We then use Herwig++ \cite{Bahr:2008pv} to generate events 
and analyse these with \Checkmate{} using the following input parameter card:
\begin{Verbatim}[commandchars=\\\@\@,frame=leftline]
## General Options
[Mandatory Parameters]
\userinputcolor Name: My_Trilepton_Run
\userinputcolor Analyses: atlas_1402_7029X

## Process Information (Each new process 'X' must start with [X])
\userinputcolor [Simplified]
\userinputcolor XSect: 2.893 FB
\userinputcolor XSectErr: 0 FB
\userinputcolor Events: /hdd/data/validation/trileptons/exclusion_A/mC1N2_625_0_mN1_375_0.hepmc
\end{Verbatim}
Looking at the \verb@/analysis/@ sub-folder within the results directory reveals the 
desired\linebreak \verb@000_atlas_1402_7029X_cutflow.dat@ and \verb@000_atlas_1402_7029X_signal.dat@ files we booked in 
our analysis code. The cutflow file tells us the number of 
events remaining after each selection step in our analysis\footnote{Note that if you do the 
same, you will  find slightly different numbers due to finite Monte Carlo statistics.}:
\begin{Verbatim}[commandchars=\\\@\@,frame=leftline]
# ATLAS
# 3 leptons
# tutorial

Inputfile:       /tutorial/results/My_Trilepton_Run/delphes/000_delphes.root
XSect:           2.893 fb
Error:           0 fb
MCEvents:        10000
 SumOfWeights:   10000
 SumOfWeights2:  10000
 NormEvents:     58.7279

Cut            Sum_W  Sum_W2  Acc     N_Norm    
CR0_All        10000  10000   1       58.7279   
CR1_Trigger    7448   7448    0.7448  43.7405   
CR2_Selection  1647   1647    0.1647  9.67249   
CR3_0Tau       1306   1306    0.1306  7.66986   
CR3_1Tau       272    272     0.0272  1.5974    
CR3_2Tau       69     69      0.0069  0.405223
\end{Verbatim}
Furthermore we can check the numbers for each signal region:
\begin{Verbatim}[commandchars=\\\@\@,frame=leftline]
# ATLAS
# 3 leptons
# tutorial

Inputfile:       /hdd/sandbox/trileptonupdate/results/My_Trilepton_Run/delphes/000_delphes.root
XSect:           2.893 fb
 Error:          0 fb
MCEvents:        10000
 SumOfWeights:   10000
 SumOfWeights2:  10000
 NormEvents:     58.7279

SR         Sum_W  Sum_W2  Acc     N_Norm      
SR0taua01  1      1       0.0001  0.00587279  
SR0taua02  4      4       0.0004  0.0234912   
SR0taua03  4      4       0.0004  0.0234912   
[...]   
SR0taub    4      4       0.0004  0.0234912   
SR1tau     20     20      0.002   0.117456    
SR2taua    38     38      0.0038  0.223166    
SR2taub    11     11      0.0011  0.0646007
\end{Verbatim}
Finally, \Checkmate{} tells us \verb@Result: Allowed, Result for r: r_max = 0.36209@. One might be 
surprised if one compares to Fig.~7a in~\cite{Aad:2014nua} in which the exact same point lies very close to 
the exclusion line. However, the discrepancy is to be expected as \Checkmate{} only considers the 
most sensitive signal region whereas within~\cite{Aad:2014nua}, sensitivities of various signal regions 
are combined using correlations to which \Checkmate{} does not have access. It is therefore reasonable 
that \Checkmate{} returns a weaker limit which is however still close to the nominal limit. 

To finish this subsection, we would like to add several general remarks on the validation 
procedure. Firstly, in most cases the results are sensitive to the actual Monte Carlo 
generator used. This is especially true if the signal regions probe events that depend 
on the additional jet radiation from the initial state. The user should make sure to use the 
same Monte Carlo and settings as those reported by the collaborations. Secondly, one should 
keep in mind that on many occasions the cutflow is given for a specific 
subprocess, e.g.\ only some of the physically allowed decays are actually included in 
the produced sample. Another possible issue is the use of truth 
level/generator cuts, e.g.\ in the cutflow sample only events with at least one truth 
lepton of $p_T > 10$~GeV are included. Since these settings are not always straightforward to 
implement, it can be convenient to start the comparison of cutflows at some 
later stage rather that at the initial 'raw' events. Similarly, users may also wish to 
produce a subset of events actually analysed by the experiment, e.g.\ only include 
leptonic $W$ events in a di-lepton analysis. This has an advantage of reducing computation time and increasing statistical accuracy. 

From a practical point of view, trivial coding errors can be spotted in the cutflow 
by noting that for some of the steps there is no change in the efficiency or the 
change is much larger than the one reported by the experiment. Generally, even when the 
numbers between \Checkmate\ and the experiment are not in perfect agreement, the general trends 
in the cutflow should remain similar. Still, one should keep in mind that Delphes is only 
a rough parametrisation of that detector response and differences of 10\%--20\% occasionally can occur. Last but not least, if the discrepancy persist, it is worth 
to directly contact the conveners responsible for the analysis that the user is trying to implement. It may well turn out that some 
details of the simulation are not explicitly mentioned in the experimental report.  

\subsection{Example 3: Inventing a completely new analysis}
So far, we have discussed the implementation of current experimental collider studies. In this section, we want to implement a completely new analysis, a monojet search at 14 TeV with an integrated luminosity of 100 fb$^{-1}$. The monojet analysis at 14 TeV is based on Ref.~\cite{Aad:2014nra} but with additional signal regions accommodating harder transverse momentum cuts on the leading jet and stricter cuts on missing transverse energy. 

We run the AnalysisManager and enter the general information as previously described. We call our analysis:

\begin{Verbatim}[commandchars=\\\@\@,frame=leftline]
    Analysis Name: 
     \userinputcolor atlas_monojet_at_14_tev 
\end{Verbatim}

We define five signal regions: 

\begin{Verbatim}[commandchars=\\\@\@,frame=leftline]
  2. Information on Signal Regions
    List all signal regions (one per line, finish with an empty line):
     \userinputcolor M1
     \userinputcolor M2
     \userinputcolor M3
     \userinputcolor M4
     \userinputcolor M5     
\end{Verbatim}

We now have to provide the background numbers for each signal region. However, these numbers are evidently not known at this stage, so we choose no.

\begin{Verbatim}[commandchars=\\\@\@,frame=leftline]
    Is the SM expectation B known? [(y)es, (n)o]?
       \userinputcolor n
\end{Verbatim}

The AnalysisManager will display the following message:

\begin{Verbatim}[frame=leftline]
Signal regions are registered but without any numbers associated to them.
IMPORTANT: The analysis will be created and can then be used like any other
           analysis. CheckMATE will skip the model exclusion tests as long as
           the expectation is not known. You can e.g. use CheckMATE on background
           samples to estimate B and dB. As soon as you know these numbers, 
           run the AnalysisManager again and use the (e)dit feature to add them.
Press key to continue!
\end{Verbatim}

We will simulate the dominant Standard Model backgrounds and enter the background numbers for all signal regions later, because for that we have to write the analysis first. In the next step we have to define all detector level objects as discussed in the previous examples.

\begin{Verbatim}[commandchars=\\\@\@,frame=leftline]
    3. Settings for Detector Simulation
        3.1: Miscellaneous
            To which experiment does the analysis correspond? [(A)TLAS, (C)MS]
               \userinputcolor A
        3.2: Electron Isolation
            Do you need any particular isolation criterion? [(y)es, (n)o]
               \userinputcolor n
        3.3: Muon Isolation
            Do you need any particular isolation criterion? [(y)es, (n)o]
               \userinputcolor y
            Isolation 1:
                Which objects should be considered for isolation? [(t)racks, (c)alo objects?]
                   \userinputcolor t
                What is the minimum pt of a surrounding object to be used for isolation? [in GeV]
                   \userinputcolor 1.0
                What is the dR used for isolation?
                   \userinputcolor 0.2
                Is there an absolute or a relative upper limit for the surrounding pt? [(a)bsolute, (r)elative]
                   \userinputcolor a
                What is the maximum surrounding pt used for isolation [in GeV]?
                   \userinputcolor 1.8
            
            Do you need more isolation criteria? [(y)es, (n)o]
               \userinputcolor n
\end{Verbatim}               
\begin{Verbatim}[commandchars=\\\@\@,frame=leftline]               
        3.4: Photon Isolation
            Do you need any particular isolation criterion? [(y)es, (n)o]
               \userinputcolor n
        3.5: Jets
            Which dR cone radius do you want to use for the FastJet algorithm?
               \userinputcolor 0.4
            What is the minimum pt of a jet? [in GeV]
               \userinputcolor 20.0
            Do you need a separate, extra type of jet? [(y)es, (n)o]
               \userinputcolor n
            Do you want to use b-tagging? [(y)es, (n)o]
               \userinputcolor n
            Do you want to use tau-tagging? [(y)es, (n)o]
                \userinputcolor n
\end{Verbatim}

After answering the remaining questions of the AnalysisManager, the source and header files are generated. Everything is now set by the AnalysisManager and we start to write the analysis code. 

We want to select events satisfying the following trigger condition and kinematic cuts. First, we apply an online trigger condition with \etmiss{}$> 80$ GeV and assume a flat trigger efficiency of 100\%. All events are required to have \etmiss{} $>150$~GeV and the leading jet must satisfy $p_T>150$ GeV and $|\eta|<2.8$. Events with isolated electrons (muons) with $p_T>20$ ($p_T>10$~GeV) are rejected. We also veto events if there are more than three jets with $p_T>30$ GeV and $|\eta|<2.8$. In order to reduce the QCD background, we require an azimuthal separation between the jets and the missing transverse momentum vector: $\Delta\phi({\rm jet},\mathbf{p}_T^{\rm miss})>0.4$. Finally, we 
choose the following set of cuts to define our signal regions.

\begin{itemize}
\item {\bf M1:} \etmiss{}$>220$ GeV and $p_T^{\rm leading jet}>280$ GeV
\item {\bf M2:} \etmiss{}$>340$ GeV and $p_T^{\rm leading jet}>340$ GeV
\item {\bf M3:} \etmiss{}$>450$ GeV and $p_T^{\rm leading jet}>450$ GeV
\item {\bf M4:} \etmiss{}$>500$ GeV and $p_T^{\rm leading jet}>500$ GeV
\item {\bf M5:} \etmiss{}$>550$ GeV and $p_T^{\rm leading jet}>550$ GeV
\end{itemize}

We now have all the information to write the analysis code. Since we have already discussed two examples, we just present the full analysis code without discussing it in detail.
\begin{Verbatim}[frame=leftline]
#include "atlas_monojet_at_14_tev.h"

void Atlas_monojet_at_14_tev::initialize() {
  setAnalysisName("atlas_monojet_at_14tev");          
  setInformation(""
  "@#targets generic ATLAS monojet studies\n"
  "@#14 TeV, 100/fb\n"
  "");
  setLuminosity(100.0*units::INVFB);      
  ignore("towers"); // These won't read tower or track information from the
  ignore("tracks"); //  Delphes output branches to save computing time.
  bookSignalRegions("M1;M2;M3;M4;M5;");
}
\end{Verbatim}
\begin{Verbatim}[commandchars=\\\@\@,frame=leftline]
void Atlas_monojet_at_14_tev::analyze() {
\userinputcolor  missingET->addMuons(muonsCombined);
\userinputcolor  electronsMedium = filterPhaseSpace(electronsMedium, 20., -2.47, 2.47);           
\userinputcolor  muonsCombined = filterPhaseSpace(muonsCombined, 10., -2.4, 2.4);         
\userinputcolor  jets = filterPhaseSpace(jets, 20., -2.8, 2.8);
  
\userinputcolor  jets = overlapRemoval(jets, electronsMedium, 0.2);
\userinputcolor  electronsMedium = overlapRemoval(electronsMedium, jets, 0.4);
\userinputcolor  muonsCombined = overlapRemoval(muonsCombined, jets, 0.4);
  
\userinputcolor  std::vector<Muon*> isoMuons = filterIsolation(muonsCombined);
  
\userinputcolor  //trigger
\userinputcolor  if (missingET->P4().Et() < 80.) 
\userinputcolor      return;
  
\userinputcolor  if (isoMuons.size() > 0 || electronsMedium.size() > 0 ) 
\userinputcolor      return;
  
\userinputcolor  if ( jets.size() > 3 && jets[3]->PT > 30. ) 
\userinputcolor      return;
  
\userinputcolor  for ( int i = 0; i < jets.size() && jets[i]->PT > 30.; i++) {
\userinputcolor    if ( fabs(missingET->P4().DeltaPhi(jets[i]->P4())) < 0.4) 
\userinputcolor      return;
\userinputcolor  }

\userinputcolor  if ( jets.size() == 0 || jets[0]->PT < 150. ) 
\userinputcolor      return;

\userinputcolor  if (missingET->P4().Et() < 150.) 
\userinputcolor      return; 

\userinputcolor  if (jets[0]->PT > 280.) {
\userinputcolor    if (missingET->P4().Et() > 220.)
\userinputcolor      countSignalEvent("M1");
\userinputcolor  }
  
\userinputcolor  if (jets[0]->PT > 340.) {
\userinputcolor    if (missingET->P4().Et() > 340.)
\userinputcolor      countSignalEvent("M2");
\userinputcolor  }
  
\userinputcolor  if (jets[0]->PT > 450.) {
\userinputcolor    if (missingET->P4().Et() > 450.)
\userinputcolor      countSignalEvent("M3");
\userinputcolor  }

\userinputcolor  if (jets[0]->PT > 500.) {
\userinputcolor    if (missingET->P4().Et() > 500.)
\userinputcolor      countSignalEvent("M4");
\userinputcolor  }

\userinputcolor  if (jets[0]->PT > 550.) {
\userinputcolor    if (missingET->P4().Et() > 550.)
\userinputcolor      countSignalEvent("M5");
\userinputcolor  }
}

void Atlas_monojet_at_14_tev::finalize() {
}       
\end{Verbatim}
After having implemented the monojet analysis code, we have to determine the expected numbers of Standard Model background events for all signal regions. Firstly, we briefly discuss the major backgrounds to the monojet signal. The $Z(\rightarrow\nu\bar\nu)+j$ is the dominant 
irreducible background. In general, $W(\rightarrow\ell\nu)+j$ is a non-negligible background even though it 
contains one charged lepton ($\ell=e^\pm,\mu^\pm)$. However, if the charged lepton is not reconstructed, it will 
have similar final states as the monojet signature. The hadronic $\tau$ decays of $W+j$ production 
also yield a sizeable contribution to the background of the monojet signal. The $t\bar t$
background happens to be relatively small and consequently we 
will omit it for simplicity as well as the other sub-dominant 
backgrounds. We have simulated the $Z+j$ and $W+j$ samples with the MC event generator Sherpa~\cite{Gleisberg:2008ta} using leading order matrix elements for up to 3 partons and using massive $b$/$c$ 
quarks with the CTEQ10 parton distribution functions~\cite{Lai:2010vv}. Our background estimate is 
given in Table~\ref{tab:monojet_background}. Now we have to provide our background numbers for each signal 
region from Table~\ref{tab:monojet_background} to \Checkmate. We again call the AnalysisManager but this time
choose the \verb@(e)dit analysis information@ option.

\begin{table}
\begin{tabularx}{\textwidth}{c || X X X X X}
\hline
Signal Region & M1 & M2 & M3 & M4 & M5 \\
SM prediction &  $733900\pm 4860$ & $197501\pm 2592$ & $51776\pm 1231$ & $18461\pm 640$ & $7477\pm 364$ \\
\hline
\end{tabularx}
\caption{Number of expected background events for all signal regions. Only statistical errors are included. All numbers are normalized to 100 fb$^{-1}$ at $\sqrt{s}=14$ TeV.}
\label{tab:monojet_background}
\end{table}

\begin{Verbatim}[frame=leftline]
   What would you like to do? 
    -(l)ist all analyses,
    -(a)dd a new analysis to CheckMATE,
    -(e)dit analysis information,
    -(r)emove an analysis from CheckMATE]
        e
    You can now edit or update some of the information for a given analysis.
    Enter the identifier of the analysis you want to edit: 
        atlas_monojet_at_14_tev
    Analysis atlas_monojet_at_14_tev successfully read in.
    Do you want to...
        ...list all defined entries? (l)
        ...restart the detector settings questions? (d)
        ...enter signal region numbers for observation and background? (n)
\end{Verbatim}

We choose \verb@n@ and enter the number of {\it observed} events  (which for a hypothetical analysis
will be the same\footnote{We implicitly assume the null hypothesis here, ie.\ that no signal above the SM expectation is present.} as the expected background), expected background events and the error of the 
background events for each signal region. The AnalysisManager will then calculate
the model independent upper limits.

We are now in the position to perform studies with our new analysis and as an example we choose to 
apply our implementation to a simple supersymmetric scenario. We consider 
direct production of scalar top quark (stop) pairs in association with one jet \cite{Drees:2012dd}. We assume that the stops decay 
into a charm quark and the lightest neutralino with a branching ratio of 100$\%$. If the 
mass splitting between the stop and the neutralino is very small, the charm jets cannot be 
reconstructed and hence our signal is a single highly energetic jet recoiling 
against missing momentum. We choose a benchmark point 
with $m_{\tilde t_1}=400$~GeV and $m_{\tilde\chi_1^0}=350$~GeV. We calculate the cross section with 
Prospino~\cite{Beenakker:1996ed} and obtain $\sigma=2.15$~pb. The events are generated as a
matched sample $p p \rightarrow \tilde t_1 \tilde t_1^*+j$ with Madgraph\_aMC@NLO~\cite{Alwall:2014hca} and the showering is 
done with Pythia~\cite{Sjostrand:2006za}. Analysing our resulting event file 
with \Checkmate, we obtain a $r$-value\footnote{The r-value is defined as the number 
of expected signal events divided by the expected 95\% C.L.\ value.} of 0.32 for M1 which
is determined to be the most sensitive signal region. Consequently we conclude that 
the parameter point is expected to be allowed in this analysis after 100~fb$^{-1}$ at LHC-14.

\subsection{Auxiliary Information}
The following functionalities might also be useful for the user in certain scenarios, which is why we list them 
separately here without reference to a particular example:

\subsubsection{File Booking}
The \Checkmate{} internal files \texttt{\_signal.dat} are automatically put into the globally chosen output directory with individual 
prefixes for each analysis and each separate input file. Sometimes a user might want to store extra information that is determined 
during the analysis, as an example kinematical distributions in order to draw histograms, in 
separate files. \Checkmate{} provides simple to use functions that create these files in the same manner as the \texttt{\_signal.dat} files, i.e. into the user--chosen output directory with individual prefixes.

To do this, the user should use the \texttt{bookFile(filename, noHeader)} function within the \texttt{initialize()} part of the analysis code (similarly to \texttt{bookSignalRegions}): this will create a file with the correct prefix before \texttt{filename} in the \Checkmate{} output directory. If one wants to create a text file and  wants general analysis information to automatically be written to the top of that file, \texttt{noHeader} should be set to \texttt{false}, otherwise \texttt{true}. The function returns an integer number, which should always be stored by the user in a variable, for example \texttt{int iFile = bookFile("hist.txt", false);}. 

Then, \texttt{fNames[iFile]} can be used to get the absolute path to the file, which can e.g.\ be used to book histograms via \Root\ (see \Root\ documentation for more information on \Root\ histogramming). 

Alternatively, one can write text information directly into the file by using \texttt{fStreams[iFile]} for example as follows: \texttt{fStreams[iFile] << "I have a jet with pt " << jet[0]->PT << std::endl;}. If \texttt{iFile} is defined as a member of the analysis class, the above code can be used in all three functions \texttt{initialize(), analyze(), finalize()}.

\subsubsection{Random Numbers}
If random numbers have to be used, \texttt{rand()/(RAND\_MAX+1.)} should be used to get a uniformly distributed random 
number between 0 and 1. The reason to use \texttt{rand()} and not e.g.\ \Root\ based random number generators is that \Checkmate{} has a 
parameter to fix the random seed of a run. This fixed seed only applies for the \texttt{rand()} function. As a fixed seed should lead 
to completely deterministic results, one should avoid using any other random number generator.

\section{Summary\label{sec:summary}}

We have introduced the AnalysisManager of \Checkmate\ that allows for the easy implementation 
of new analyses, either invented by user or following experimental searches. The program guides the user through the numerous choices possible for final 
state objects, simplifying many of the complicated LHC definitions. In order to aid the actual coding of
analyses, many useful,  routinely performed functions are included. Additionally, a suite 
of common LHC kinematical variables are also implemented. Finally, all the statistical variables already
available in \Checkmate\ can be calculated automatically. \\

\Checkmate\ can be downloaded from:

\url{http://checkmate.hepforge.org/} \\

Detailed program documentation can be found at:

\url{http://checkmate.hepforge.org/documentation/index.html}

\section*{Acknowledgements}

We would all like to thank Sabine Kraml and LPSC Grenoble for support and hospitality while part of this 
manuscript was prepared. We would also like to thank Liang Shang for the careful reading of our manuscript. J.T. would
like to thank Herbi Dreiner and Bonn University for additional kind support. 
The work of J.S. Kim and K. Rolbiecki has been partially supported by the MINECO, Spain, under contract FPA2013-44773-P; 
Consolider-Ingenio CPAN CSD2007-00042 
and  the Spanish MINECO Centro de excelencia Severo Ochoa Program under grant SEV-2012-0249. K. Rolbiecki was also supported by JAE-Doc programme. 




\bibliographystyle{JHEP}
\bibliography{manual_analysis}

\end{document}